\documentclass[iop]{emulateapj}

\newcommand{\nstars}{\hbox{120 }}

\newcounter{minirefcount}
\usepackage{amssymb}
\usepackage{amsmath}
\usepackage{wasysym}
\newcommand{\mr}[2]{\refstepcounter{minirefcount}\label{#2}(\arabic{minirefcount}) #1}

\shorttitle{Measuring Disk Accretion with H I Pf$\beta$}
\shortauthors{Salyk et al.}

\begin{document}
\title{Measuring Protoplanetary Disk Accretion with H I Pfund $\beta$ $^1$}  
\author{Colette Salyk $^2$}
\affil{National Optical Astronomy Observatory, 950 N Cherry Ave, Tucson, AZ 85719, USA}
\email{$^1$Accepted for publication in the Astrophysical Journal}
\email{$^2$csalyk@noao.edu}
\author{Gregory J.\ Herczeg}
\affil{The Kavli Institute for Astronomy and Astrophysics at Peking University, Yi He Yuan Lu 5, Hai Dian Qu, Beijing 100871, P. R. China }
\author{Joanna M.\ Brown}
\affil{Harvard-Smithsonian Center for Astrophysics, 60 Garden Street, Cambridge, MA 02138, USA}
\author{Geoffrey A.\ Blake}
\affil{Division of Geological \& Planetary Sciences, Mail Code 150-21, California Institute of Technology, Pasadena, CA 91125, USA}
\author{Klaus M.\ Pontoppidan}
\affil{Space Telescope Science Institute, 3700 San Martin Drive, Baltimore, MD 21218, USA}
\author{Ewine F.\ van Dishoeck}
\affil{Leiden Observatory, Leiden University, P.O. Box 9513, 2300 RA Leiden, the Netherlands}
\affil{Max-Planck-Institut f\"{u}r Extraterrestrische Physik, Giessenbachstrasse 1, 85748 Garching, Germany}

\begin{abstract}

In this work, we introduce the use of H \textsc{i} Pfund $\beta$ (Pf$\beta$; 4.6538 $\mu$m) as a tracer of mass accretion from protoplanetary disks onto young stars.  Pf$\beta$ was serendipitously observed in NIRSPEC and CRIRES surveys of CO fundamental emission, amounting to a sample size of \nstars young stars with detected Pf$\beta$ emission.  Using a subsample of disks with previously measured accretion luminosities,  we show that Pf$\beta$ line luminosity is well correlated with accretion luminosity over a range of at least three orders of magnitude.  We use this correlation to derive accretion luminosities for all \nstars targets, 65 of which are previously unreported in the literature.  The conversion from accretion luminosity to accretion rate is limited by the availability of stellar mass and radius measurements; nevertheless, we also report accretion rates for 67 targets, 16 previously unmeasured. Our large sample size and our ability to probe high extinction values allow for relatively unbiased comparisons between different types of disks.  We find that the transitional disks in our sample have lower than average Pf$\beta$ line luminosities, and thus accretion luminosities, at a marginally significant level. We also show that high Pf$\beta$ equivalent width is a signature of transitional disks with high inner disk gas/dust ratios.  In contrast, we find that disks with signatures of slow disk winds have Pf$\beta$ luminosities comparable to those of other disks in our sample.   Finally, we investigate accretion rates for stage I disks, including significantly embedded targets.  We find that stage I and stage II disks have statistically indistinguishable Pf$\beta$ line luminosities, implying similar accretion rates, and that the accretion rates of stage I disks are too low to be consistent with quiescent accretion.  Our results are instead consistent with both observational and theoretical evidence that stage I objects experience episodic, rather than quiescent, accretion.

\end{abstract}
 

\section{INTRODUCTION}
In the study of protoplanetary disks and protostars, much effort has been focused on the study of mass accretion rates --- the rates at which mass is transferred from circumstellar disks to stars --- because it is so intricately linked to processes important for star and planet formation.  Mass accretion is a measure of the viscosity of the disk and determines the overall rate of mass and momentum transfer, and thus the pace of disk evolution.  The rate of mass accretion will affect the disk lifetime (and thus the time available for planet formation) and the rate of planetary migration, and may in turn be a {\it tracer} of the presence of planets \citep[e.g.,][]{Alexander07}.  Accretion greatly affects the inner disk environment, with the disk truncated and material lofted onto the star at or near the stellar corotation radius \citep{Shu94}, which could be related to the observed pile-up of giant planets at small orbital radii \citep{Lin96, Butler06}.  Finally, accretion is a tracer of star/disk magnetic interactions and determines the early angular momentum evolution of the star \citep[e.g.,][]{Agapitou00}. 

The most accurate measurement of mass accretion is likely obtained from the spectroscopic observation and modeling of UV excess flux \citep[e.g.,][]{Valenti93, Gullbring98, Herczeg08}, as this provides a nearly direct measure of the total accretion luminosity.  However, there has long been an interest in measuring accretion rates with data that are relatively simpler to obtain and analyze, as well as data that can be obtained at longer wavelengths, where extinction is lower.  This has led to a number of studies of easily-observable H \textsc{i} emission lines believed to be produced in the accretion columns and accretion shock along with the UV continuum excess \citep{Calvet98}.  Such studies have shown that H \textsc{i} line luminosities correlate with UV-excess-derived accretion luminosities, and thus can be used as reasonably reliable tracers of mass accretion rates \citep[e.g.,][]{Muzerolle98c,Natta04, GarciaLopez06, Natta06, Fang09}, albeit with some systematic uncertainties and caveats \citep[e.g.,][]{Herczeg08}.  The ease of collecting spectra of H \textsc{i} emission lines results in more comprehensive samples of mass accretion rates, allows for the study of accretion rates in more embedded disks, and allows for a simultaneous measure of accretion rates and other disk properties, such as veiling by the disk continuum, or molecular emission line strengths.  In fact, one of the best studied H \textsc{i} lines --- H$\alpha$ --- remains a popular means of estimating mass accretion, in spite of the fact that models suggest it saturates even at moderate accretion rates \citep{Muzerolle98a}, and the fact that its strength and profile shape can also be affected by other parameters, including outflow rate and system inclination \citep{Kurosawa06}.  In addition, since measured accretion rates span several orders of magnitude, even measurements with error bars as large as $\sim 0.5-1$ dex can be used to broadly characterize a sample of objects and provide scientifically useful information.

In this work, we introduce a new tracer of mass accretion --- specifically, H \textsc{i} Pf$\beta$ --- that offers several advantages over other tracers, and results in one of the most comprehensive coherent datasets of a single mass accretion tracer to date.  H \textsc{i} Pf$\beta$ ($n=7\rightarrow5$; hereafter Pf$\beta$) is located at 4.6538 $\mu$m in the M band, which places it between the R(0) and R(1) lines of the CO ro-vibrational fundamental band.  Thanks to two large campaigns designed to study CO fundamental emission in protoplanetary disks with Keck-NIRSPEC \citep[e.g.,][]{Blake04, Salyk11b} and VLT-CRIRES \citep[e.g.,][]{Pontoppidan11a, Herczeg11, Brown13}, Pf$\beta$ has been serendipitously observed in more than 100 young stars with disks.  Pf$\beta$ has similar equivalent widths to other infrared tracers, including Br$\gamma$ (n=7$\rightarrow$4), with which it shares the same upper level energy, and appears to be ubiquitous for disks determined by other means to be typical active accretors.  It also offers several distinct advantages over other accretion tracers due to its position in the M band. Firstly, it is an ideal tracer of accretion rates in transitional disks, as the low M-band photospheric flux results in a very high Pf$\beta$ equivalent width when the dust continuum is low, as it is in transitional disks.  Secondly, the high degree of veiling in the M band relative to shorter wavelengths makes contamination from stellar photospheric absorption negligible in nearly all cases. Thirdly, tentative discoveries of young planets in disks \citep{Huelamo11, Kraus12} have sparked interest in relating CO emission line variability to the tidal influence of proto-planets \citep{Regaly11}.  An understanding of the connection between accretion and Pf$\beta$ may help observers differentiate between accretion- and planet-induced variability observed in CO emission lines.  Finally, Pf$\beta$ is observable even in heavily extincted disks, and so can be used to study the youngest disks embedded in their natal cloud.  

The ability to measure accretion rates in embedded disks is a particularly important strength of Pf$\beta$ as an emission tracer.  A star is thought to gain most of its mass while it is still embedded in a dense envelope, prompting the naive expectation that accretion rates should be higher during these stages.  During this phase, however, the commonly-used accretion indicators in the optical and near-IR are not usually observable and this expectation is difficult to confirm.  As a substitute for direct measurements of accretion rates, the mass history of stars has been inferred from a global analysis of temperature-luminosity diagrams \citep{Kenyon90,Dunham10,Zhu10b}.  In these analyses, the luminosity distribution is $\sim 10\%$ of the expected luminosity distribution required to build stars via steady accretion, suggesting that the star may need to grow mostly in large, short bursts.  A few studies of  optical and near-IR accretion tracers in more extincted disks \citep{Muzerolle98c, White04, White07} have suggested a similar result --- namely, that accretion rates in younger disks are lower than expected.  However, as even near-IR accretion tracers like H$\alpha$, Br$\gamma$ and Pa$\beta$ are not visible in the most embedded disks, these studies may have been biased towards older systems, including possibly edge-on, evolutionarily older (yet observationally extincted) disks.  Therefore, the study of Pf$\beta$ provides an exciting new path for measuring accretion in the youngest, most embedded disks.

In this paper, we report the detection and measurement of Pf$\beta$ emission lines observed with Keck-NIRSPEC and VLT-CRIRES.  We develop a method to use Pf$\beta$ emission to measure accretion luminosity by correlating line luminosities to known accretion luminosities for a sample of young stars.  This correlation is then applied to a large sample of targets to provide accretion luminosities for \nstars young stars, including a sample of objects still embedded in their molecular envelopes.  In Section \ref{section:obs}, we briefly describe the NIRSPEC and CRIRES observations and data reduction procedures, as well as the sample selection.  In Section \ref{section:extraction}, we discuss the flux extraction procedure and corresponding uncertainties.  In Section \ref{section:correlation}, we calculate a relationship between Pf$\beta$ line luminosity and accretion rate, and in Section \ref{section:accretion} we apply this to our full sample to provide accretion luminosity estimates for \nstars stars.  In Section \ref{section:discussion}, we discuss some implications of our results.

\section{OBSERVATIONS AND REDUCTION}
\label{section:obs}
\subsection{Observing Procedures}

NIRSPEC observations were obtained with the Keck II telescope as part of a large survey of CO rovibrational emission from young stars with disks \citep[see, e.g.,][]{Salyk11b}, spanning 2001--2011.    Spectra were obtained in the high-resolution mode with R$\sim$25000, in the M-wide filter.  Echelle and cross-disperser positions were optimized for the observation of the P branch of CO v$=1\rightarrow0$.  Although the favoured echelle settings evolved somewhat throughout the course of the survey, nearly all sources were observed in a setting that included the H \textsc{i} lines Pf$\beta$  (4.6538 $\mu$m) and Hu$\epsilon$ (4.6725 $\mu$m).  NIRSPEC targets were observed in ABBA nod sets, with AB pairs differenced to remove thermal emission from Earth's atmosphere.
Integration times (exposure time multiplied by coadds) in each position were limited to 1 minute to minimize atmospheric variation between frames of a pair.
To correct for telluric absorption, A or B telluric standard stars with nearly blackbody spectra were observed close in time and airmass to the targets. 

CRIRES observations were obtained with the VLT as part of a large survey of CO rovibrational emission from young stars with circumstellar disks \citep{Pontoppidan11a, Brown13}, including a significant sample of embedded protostars \citep{Herczeg11}.  Only $\sim$0.02 $\mu$m are covered in each integration with the non-cross-dispersing CRIRES instrument, but the majority of targets in this survey were still observed in a setting that included Pf$\beta$.  Observations were obtained with the 0''.2 slit, resulting in a resolution of $R\sim$90,000.  Targets were observed in ABBA nod sets, with 60 second integration times for each image.  Random dithers between 0 and 1'' were added to the standard 10'' nod sequence to distribute the counts over different pixels, and ensure that the data would not always land on a bad pixel.  When possible, brighter targets were observed with adaptive optics.  

A log of the NIRSPEC and CRIRES observations included in this study can be found in Table \ref{table:obslog}.

\subsection{Data Reduction}
NIRSPEC data reduction routines were developed by our team \citep{Blake04} in IDL.  NIRSPEC spectra were first linearized using the shape of telluric emission lines and the observed spectra.  Fluxes were extracted with a 2.8$\sigma$ aperture around the best-fit PSF center, and nearby columns were subtracted to account for any residual sky emission in the A--B difference image.  Extracted spectra were wavelength calibrated utilizing telluric lines.  Telluric standards were processed with the same procedures as the target stars.  Target spectra were then divided by the telluric standard star spectra (adjusted for small differences in airmass) to obtain spectra corrected for telluric absorption.  

\begin{figure}
\epsscale{1.1}
\plotone{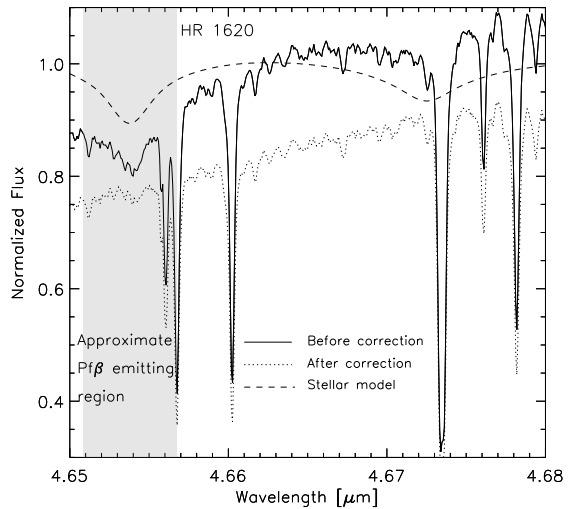}
\caption{Solid and dotted lines show sample standard-star spectrum before and after stellar atmospheric model division.  The shaded region marks an approximate location for Pf$\beta$ emission, assuming
a width of 190 km s$^{-1}$.  The actual region will depend on the line width and relative radial velocity between the source and the standard stars.
\label{fig:hremoval} }
\end{figure}

Although telluric standards were nearly featureless, most A stars have photospheric H \textsc{i} absorption, including that of Pf$\beta$.  Thus, correcting for this feature is crucial for obtaining accurate Pf$\beta$ fluxes in our targets.   To account for this, telluric standard spectra were first flattened utilizing Kurucz models, broadened to match observed H \textsc{i} absorption shapes.  An example of this process is shown in Figure \ref{fig:hremoval}.  Note that the corrected telluric spectrum should be linear (except for the narrow telluric absorption lines), and that this can easily be inspected by eye to confirm proper correction of the photospheric H I.  In rare cases, the H \textsc{i} absorption profiles were not well accounted for by Kurucz models; in these cases, we simply allowed for an additional Gaussian-shaped absorption component to be removed from the spectrum.  For all spectra, we performed an additional check on potential contamination from the standard star: we divided a flat standard star observation by each Kurucz-model-corrected standard star observation to confirm that it did not produce a spurious H \textsc{i} emission line.

Since the primary observational target of these surveys, CO, can overlie telluric CO absorption lines, targets were typically observed on two or more dates, and spectra combined to create complete CO line profiles.  The H \textsc{i} line emitting regions are relatively unaffected by telluric absorption and so the repeated observations simply increase the S/N in these lines.  Alternatively, multiple epochs can also be left separate to investigate accretion variability. This aspect of the dataset is left as future work, and will likely benefit from the inclusion of additional datasets from the NIRSPEC and CRIRES archives.

Approximate flux calibration was achieved by comparing telluric standard star spectra with literature photometry to derive a conversion from counts to flux.  In all cases, the brightest observation of standard and target were utilized, as these would represent the best-centered observations.  However, there is no guarantee that the sources were well centered, and the flux correction factor due to slit losses can be either greater than or less than one, depending on whether it is the source or the telluric standard that is observed off center.  Nevertheless, we find decent agreement between absolute flux and literature photometry; the RMS difference between measured and literature fluxes (see Table \ref{table:photometry}) is 50\% for our sample as a whole.

Due to the inherent difficulties with absolute flux calibration for spectroscopic observations, we have chosen to correct observed fluxes with literature photometry whenever possible.  The photometry used in this work can be found in Table \ref{table:photometry}.  We used M-band photometry when available, or secondarily an interpolation of other measurements, often Spitzer IRAC 4.5 and 5.8 $\mu$m fluxes.  Although young stars with disks are known to have variable infrared emission, the variability is typically less than 10\% for class 0-II disks \citep{Luhman10}.  Variability may be higher for transitional disks \citep[$\sim$30-50\%;][]{Espaillat11} and outbursting sources like EX Lup \citep[e.g.,][]{Aspin10}, and so we note that such sources can have a corresponding systematic error in their Pf$\beta$ line flux measurement.  If photometric measurements were not available in the literature, we simply used our own absolute photometry derived from the spectroscopic observations.

CRIRES reductions utilize routines written by K. Pontoppidan and described in \citet{Pontoppidan08}.  2D images for each source were added together after aligning and resampling onto a 2$\times$ finer grid.  Spectra were extracted using an aperture twice the size of the FWHM of the spectral profile.  As with the NIRSPEC observations, nearby telluric standards were observed close in time and airmass.  Wavelength calibration was performed on the telluric standards using telluric emission lines and was applied to the target spectra after small shifts to align the spectra of target and standard.  

In the nominal CRIRES reduction routines, target spectra were then simply divided by the telluric spectra to remove telluric absorption features.  However, this does not account for H \textsc{i} absorption in the standard stars, and would have produced anomalously strong H \textsc{i} emission in our targets.  Therefore, we implemented a routine adapted from the NIRSPEC reduction routines, which uses Kurucz photospheric models to fit and correct for H \textsc{i} absorption features in the standard stars.  

Spectra of standard stars were also compared with literature photometry to obtain a conversion from counts to pixel, which were applied to the target spectra to obtain an approximate absolute flux calibration.  However, just as with the NIRSPEC data, CRIRES flux calibration suffers from an uncertain degree of slit losses for any given observation.  Therefore, as with the NIRSPEC observations, we adjusted the spectra to match literature photometry whenever possible.  We find a few large discrepancies (factors of $\sim2-3$ between literature and measured photometry, including for Elias 29, IRS 44, LLN 8, LLN 17 and WL 12.  Since all show higher literature fluxes than CRIRES-measured fluxes, it is possible that this is because prior observations with the Spitzer Space Telescope or ISO included more than one source in their relatively large spatial point-spread-function.  For the non-outliers, we find the RMS difference between observed and literature fluxes to be $\sim$52\%.  

\subsection{Sample Selection}
The NIRSPEC and CRIRES target lists used in this work were compiled for prior studies of M-band CO emission and do not represent an unbiased sample of young stars.  However, the large sizes of these surveys means they come close to representing a complete flux-limited sample of disks with detectable CO fundamental emission in nearby clouds.  The NIRSPEC M-band survey as a whole is dominated by revealed, optically thick disks, including significant numbers of both low-mass (T Tauri) and mid-mass (Herbig Ae/Be) disks.  It also includes a smaller but significant number of embedded disks.  The NIRSPEC sample is strongly biased against sources with tenuous disks or no disks at all, as these disks tended not to produce observable CO rovibrational emission lines.  The NIRSPEC sample is also biased towards clouds at the higher declinations observable from Mauna Kea, sampling well in Taurus, Perseus, Serpens and Ophiuchus, only poorly in Lupus, and not at all at lower declinations.  The NIRSPEC sample was also limited to targets with M-band continuum fluxes brighter than $\sim$0.01 Jy (M$\sim$9).

The CRIRES M-band CO survey contains many of the bright Class II disks visible from the southern sky as well as a number of Class I embedded protostars and transitional disks. This includes targets in Lupus, Vela and Chamaeleon that are difficult or impossible to observe with NIRSPEC.

A more detailed discussion of the evolutionary status of our sample as a whole is included in Section \ref{sec:embedded}.

\section{General characteristics of Pf$\beta$ detections and non-detections}
\label{sec:characteristics}
\begin{figure}
\epsscale{1.3}
\plotone{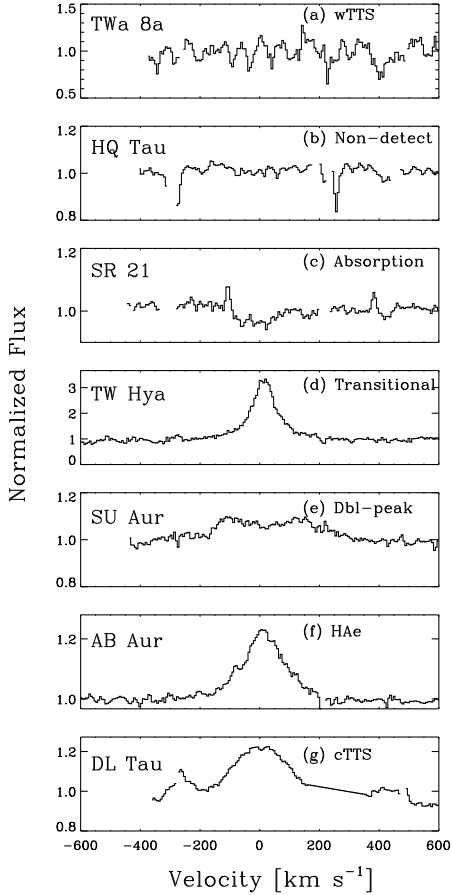}
\caption{Sample set of Pf$\beta$ emission line profiles from full NIRSPEC sample.  (Note that any overlying CO emission or absorption lines have already been removed.)
\label{fig:nirspec_profiles} }
\end{figure}

\begin{figure}
\epsscale{1.3}
\plotone{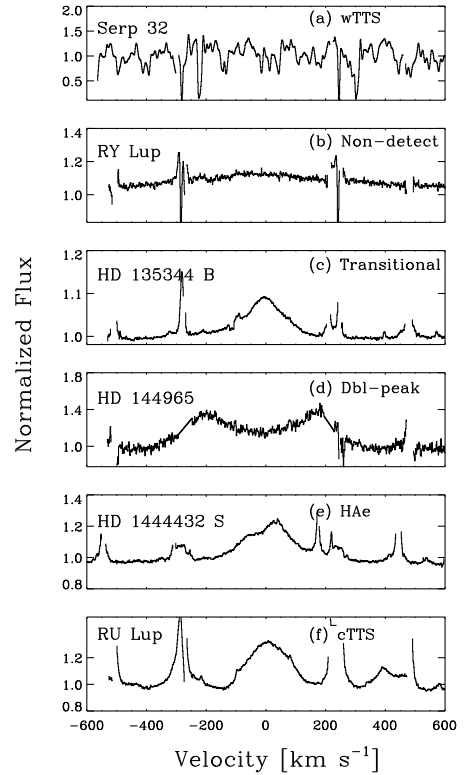}
\caption{Sample set of profiles from full CRIRES sample.  (Note that any overlying CO emission lines have already been removed.)
\label{fig:crires_profiles} }
\end{figure}

A sample of targets showing different strengths and types of Pf$\beta$ emission profiles is shown for NIRSPEC in Figure \ref{fig:nirspec_profiles} and for CRIRES in Figure \ref{fig:crires_profiles}.  Although the biases in our sample selection make any statistics difficult to interpret, Pf$\beta$ appears to be a robust tracer of accreting systems.  Greater than 80\% of optically thick, classical disks around T Tauri and Herbig Ae/Be stars show detectable Pf$\beta$ emission.  It is interesting to note that we find a slightly lower detection fraction for embedded protostars ($\sim 60$\%).  A similar fraction of embedded sources without Pf $\beta$ emission is found in a lower resolution VLT-ISAAC study of $\sim$30 sources by Pontoppidan et al. (2003, their Figures 1--7).
Although the difference in detection rates may not be significant, especially because of possible biases in the selection of embedded objects, the presence of a large number of embedded protostars with little or no detectable Pf$\beta$ is interesting in and of itself.  Because embedded protostars have rising continua towards the infrared, one might imagine that this could decrease line/continuum ratios and introduce an obsevational bias.  However, the embedded protostars with Pf$\beta$ detections do not appear to be systematically fainter than those without, nor do the embedded protostars appear to have systematically higher M-band fluxes than the protstars with classical disks.   

The majority of targets with Pf$\beta$ non-detections represent a few distinct types of objects.  Not surprisingly, debris disks and disks classified as weak-line T Tauri stars (wTTS's) or protostars with class III spectral energy distributions (SEDs) generally do not appear to show Pf$\beta$ in emission. wTTS's with weak or non-existent Pf$\beta$ emission include HD98800 (also a circumbinary disk; \citealp{Furlan06}), Hen 3-600A, LkH$\alpha$ 332 G1, TWA 7, TWA 8a and WaOph 4.  These additionally show no CO fundamental emission and very low levels of veiling, likely reflecting the expected relationship between warm inner disk gas, inner disk dust, and accretion.   Other targets without detectable Pf$\beta$ emission are in some way straddling the evolutionary boundary between optically thick and tenuous disks. These include HD 36917 (classified as transitioning between classes II and III in \citealp{Manoj02}), CoKu Tau/3 (class II in \citealt{Andrews05}; wTTS in \citealt{Furlan11}), FN Tau (classified as bordering between classical T Tauri star --- cTTS --- and wTTS in \citealt{Furlan11}), HQ Tau (wTTS by H$\alpha$ equivalent width definitions, but cTTS according to its H$\alpha$ 10\% width \citealp{Furlan11}) and WSB 60 (classified as Class II in \citealt{Evans09} but shows transitional disk millimeter emission shape in \citealt{Andrews11}).  Other targets with no detectable Pf$\beta$ emission include the transitional disks HD 149914 (which also shows no detectable Br$\gamma$ emission; \citealp{Brittain07}) as well as SR 21 \citep{Brown07}.  SR 21 has detectable CO fundamental emission, although in contrast to many other transitional disks, the emission arises from moderately large disk radii ($\sim$ 5--8 AU;  \citealp{Salyk11b, Pontoppidan11b}).  Two circumbinary disks, CoKu Tau/4 (7.8 AU; \citealp{Ireland08}) and ROXs42c (23 AU; \citealp{Kraus11}), also do not show Pf$\beta$ emission, although it should be noted that this does not appear to be universally true, as close binary DP Tau (projected separation 15.5 AU \citealp{Kraus11}) has detectable Pf$\beta$ emission.  A final interesting set of objects that universally shows no detectable Pf$\beta$ emission is the set of FU Orionis stars.  

It is interesting to ask whether any classical, optically thick disks show no detectable Pf$\beta$ emission, either to test whether Pf$\beta$ could be unreliable as an accretion tracer, or to search for unique targets with optically thick disks but no accretion.  The list of disks with class II SEDs that do not show Pf$\beta$ emission above detectable limits include c2dJ033035.92+303024.4, HK Tau, IRAS 03380+3135, IRAS 04385+2550, LkH$\alpha$ 270, LkCa 8, LkH$\alpha$ 325 and WaOph 5.  The first five show no apparent CO emission, while the final three targets show weak CO fundamental emission, and their spectra suggest the possible presence of Pf$\beta$ at low S/N; therefore, these suggest the presence of inner disk gas.  Few disks are seen with strong CO fundamental emission and no detectable Pf$\beta$ emission, one exception being the transitional disk SR 21.  Very low accretion rates are also apparently not detectable in our sample.  For example, the spectroscopic binary Hen 3-600A,  accreting at a rate of $\sim5\times10^{-11}M_\odot \mathrm{yr}^{-1}$ \citep{Muzerolle00}, does not have detectable Pf$\beta$ emission.

We investigated but did not determine any relationship between line shapes or widths and any other stellar or disk parameters, except that sources with double-peaked line profiles (including CI Tau, HD 141569 A, RY Tau and SU Aur) tend to have moderately high disk inclinations (typically 50-60$^\circ$).  A full discussion of the relationship between CO vibrational emission and Pf$\beta$ emission is left as future work.  Here we focus on targets with detectable Pf$\beta$ emission, in order to measure mass accretion rates.  Therefore, for this study we have selected a subset of the full NIRSPEC and CRIRES samples via visual confirmation of the presence of Pf$\beta$ emission.  The complete target list used in this study, with relevant stellar parameters, is shown in Table \ref{table:params}.  

Pf$\beta$ lines are unresolved in all natural-seeing NIRSPEC observations, constraining the emission to radii less than $\sim$50 AU for typical seeing and stellar distances.  Pf$\beta$ observations are also unresolved in the adaptive optics corrected CRIRES spectra. With a typical point spread function core of 0.18'', this constrains the Pf$\beta$ to radii less than $\sim$ 13 AU at 140 pc.  This is consistent with an accretion origin for Pf$\beta$, as opposed to an outflow origin.

\section{DERIVATION OF Pf$\beta$ LINE LUMINOSITIES}
\label{section:extraction}

\subsection{CO Emission Corrections}
\begin{figure*}
%\epsscale{1.1}
\plotone{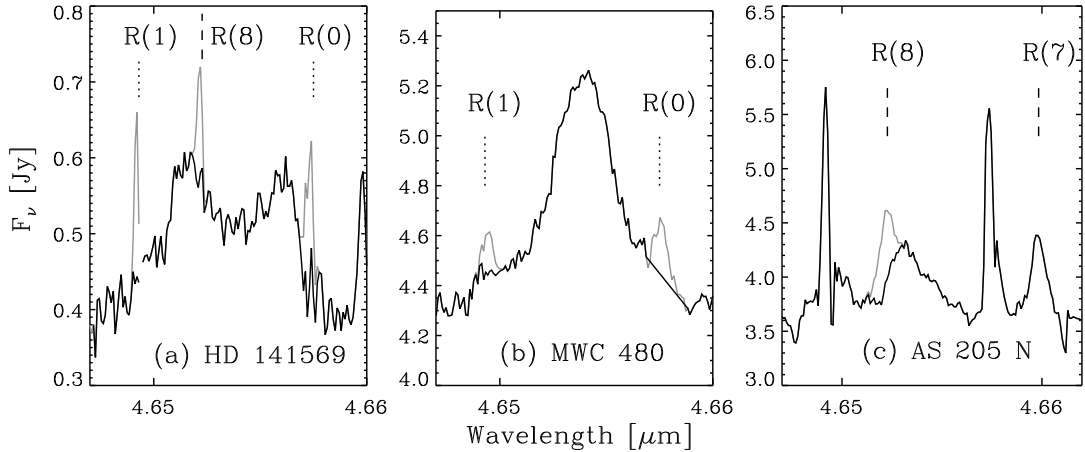}
\caption{Gray and black lines show sample spectra before and after removal of CO emission lines, respectively.  Dotted lines mark CO $1\rightarrow0$ lines; dashed lines mark CO $2\rightarrow1$ lines.  (a) Lines are removed with Gaussian fits.  (b) R(0) is removed with a linear fit to the underlying spectrum.  (c) R(8) is removed by using a Gaussian fit to R(7).
\label{fig:coremoval} }
\end{figure*}

The Pf$\beta$ line center is close to a number of CO vibrational lines, including $v=1\rightarrow0$ R(0) (4.6575 $\mu$m) and R(1) (4.6493 $\mu$m), $v=2\rightarrow1$ R(8) (4.6523 $\mu$m) and $^{13}$CO R(14) (4.6572 $\mu$m) and R(15) (4.6504 $\mu$m).  A majority of Pf$\beta$ profiles are affected by at least one of these CO features, which contain non-negligible amounts of flux.  A number of techniques were utilized to correct for the CO emission, as demonstrated in Figure \ref{fig:coremoval}.  If possible, CO lines were removed via subtraction of a Gaussian fit to the CO emission lines.  In cases where the fits were not reliable, the line was replaced by a linear interpolation of the surrounding flux.  In some cases, the nearby $v=2\rightarrow1$ or $^{13}$CO lines contributed so significantly to the flux that it was difficult to separate the CO and Pf$\beta$ contributions.  In such cases, an adjacent CO line would be shifted to the contaminating line location and subtracted.  This necessarily introduces some error (of order 10\%) due to the fact that the flux in adjacent CO lines may not be identical.  However, more precise modeling to derive these CO line fluxes was not pursued, due to severe blending of the $v=2\rightarrow1$  and $^{13}$CO rotational ladders with the much stronger $v=1\rightarrow0$ emission lines.

\subsection{Flux Extraction}
\begin{figure}
%\epsscale{0.7}
\plotone{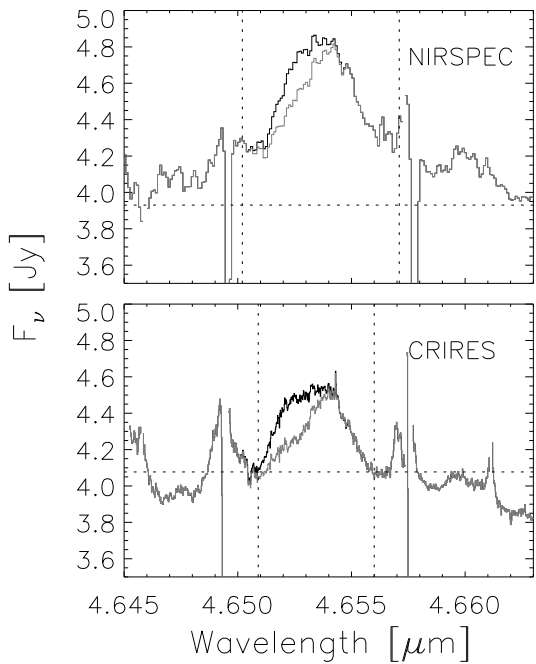}
\caption{Pf$\beta$ line profiles from HL Tau observed by NIRSPEC and CRIRES before (black) and after (gray) CO emission removal.  Dotted lines mark the regions chosen for flux extraction, demonstrating how uncertainty in the continuum level and line width can in extreme cases result in a factor of three difference in extracted line flux.
\label{fig:hltau} }
\end{figure}

Observed Pf$\beta$ line profiles are typically not described by a Gaussian or other simple functional form, and so we chose to extract fluxes by summing all flux within a defined window, after subtraction of the continuum.  The extraction window and continuum level were both determined by eye, and have associated systematic uncertainties.  These can be particularly high if a number of adjacent CO lines makes the continuum hard to define, and/or if the Pf$\beta$ profile has broad wings.  The 
sample overlap between NIRSPEC and CRIRES allows us to estimate this uncertainty to some degree, by comparing the fluxes derived from each dataset.  We find good agreement between extractions using the two datasets, with typical differences less than a factor of two.  Note that since the Pf$\beta$ line luminosity or the underlying continuum can be variable, some of this difference could be attributable to real changes that occurred between the two sets of observations. In our most egregious case, HL Tau, we find a factor of 3 discrepancy between the two datasets.  This extreme case is illustrated in Figure \ref{fig:hltau}, showing how uncertainty in both the continuum position and the position of the Pf$\beta$ line wings results in large flux differences between the two extractions.

\begin{figure*}
\includegraphics[angle=90, scale=0.8]{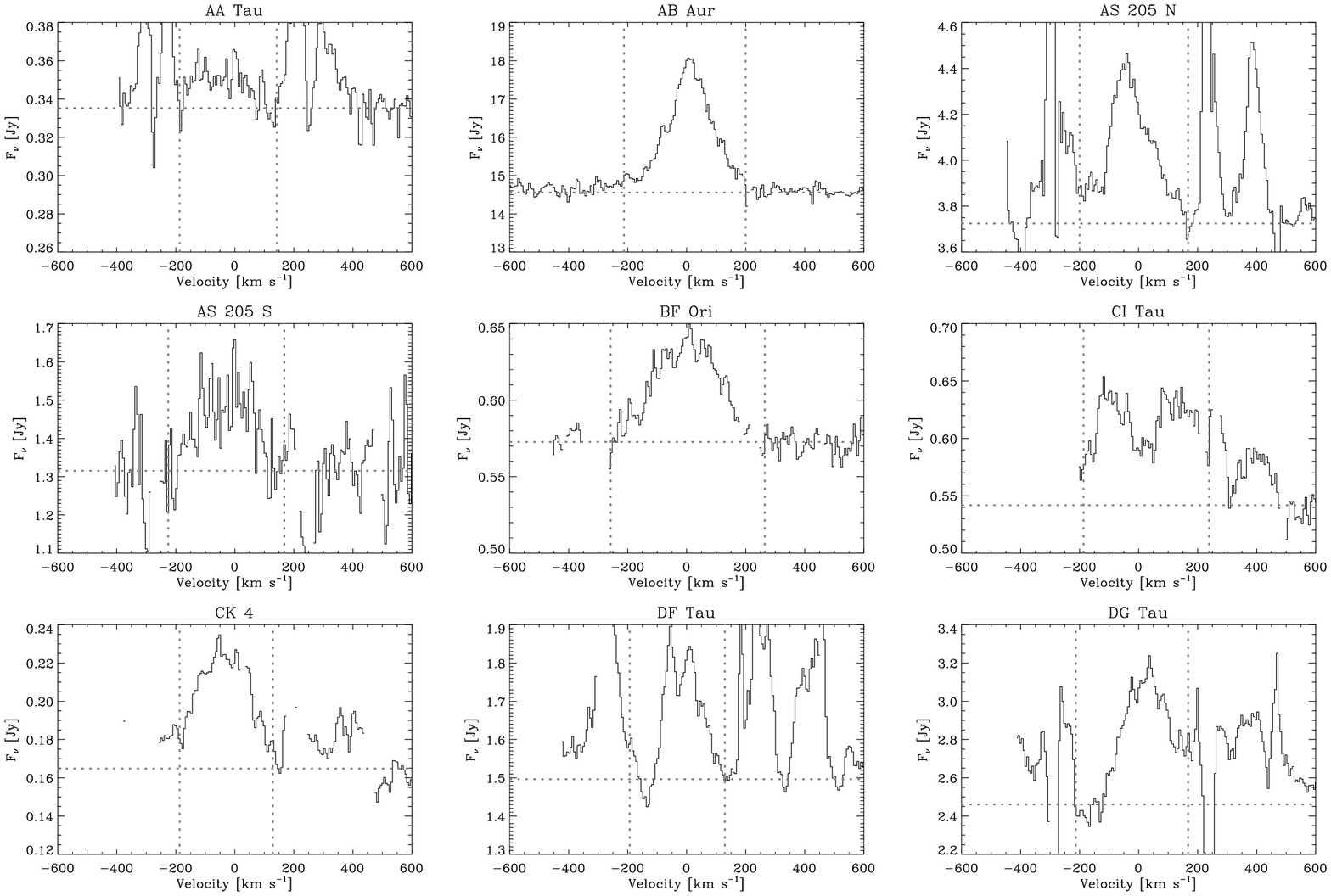}
\caption{Pf$\beta$ line profiles observed with NIRSPEC, after correction for CO emission and literature photometry, when available.   Dotted lines mark the baseline and limits used for flux extraction.  The full set of NIRSPEC and CRIRES line profiles is available in an online version of this figure.  \label{fig:hlines_paper}}
\end{figure*}

A few of the photometrically-corrected NIRSPEC line profiles are shown in Figure \ref{fig:hlines_paper};  the remaining NIRSPEC and CRIRES profiles are available in the on-line version of the article.  Our derived Pf$\beta$ luminosities, computed using the distances in Table \ref{table:params}, are listed in Table \ref{table:pf}.  Line luminosities have also been corrected for extinction, assuming an extinction of $0.034\times A_V$ \citep{Cardelli89} at 4.6538 $\mu$m, and the visual extinctions in Table \ref{table:params}; if the extinction is not known, we do not include any correction.  Note that even the maximum extinction in our sample ($A_V = 34$) only results in a change in line flux by a factor of $\sim$ 3, and thus computation of Pf$\beta$ luminosity is relatively insensitive to extinction.  When both NIRSPEC and CRIRES data were available, we show the average of those measurements.

\subsection{Importance of Photospheric H I Absorption}
\label{sec:photosphere}
While accretion flows are believed to be the source of the Pf$\beta$ emission, the observed spectra also include a contribution from the underlying stellar photosphere of the star/disk system. As we will discuss in this section, the correction for the underlying stellar photosphere should in nearly all cases be insignificant compared to other uncertainties, and we therefore do not apply any correction to most sources. (Note that this correction is distinct from the correction for Pf$\beta$ absorption in the photosphere of the telluric standard star, which is always corrected for in the data reduction process).

\begin{figure*}
 \begin{minipage}{\textwidth}
%\epsscale{0.5}
\centering
\includegraphics{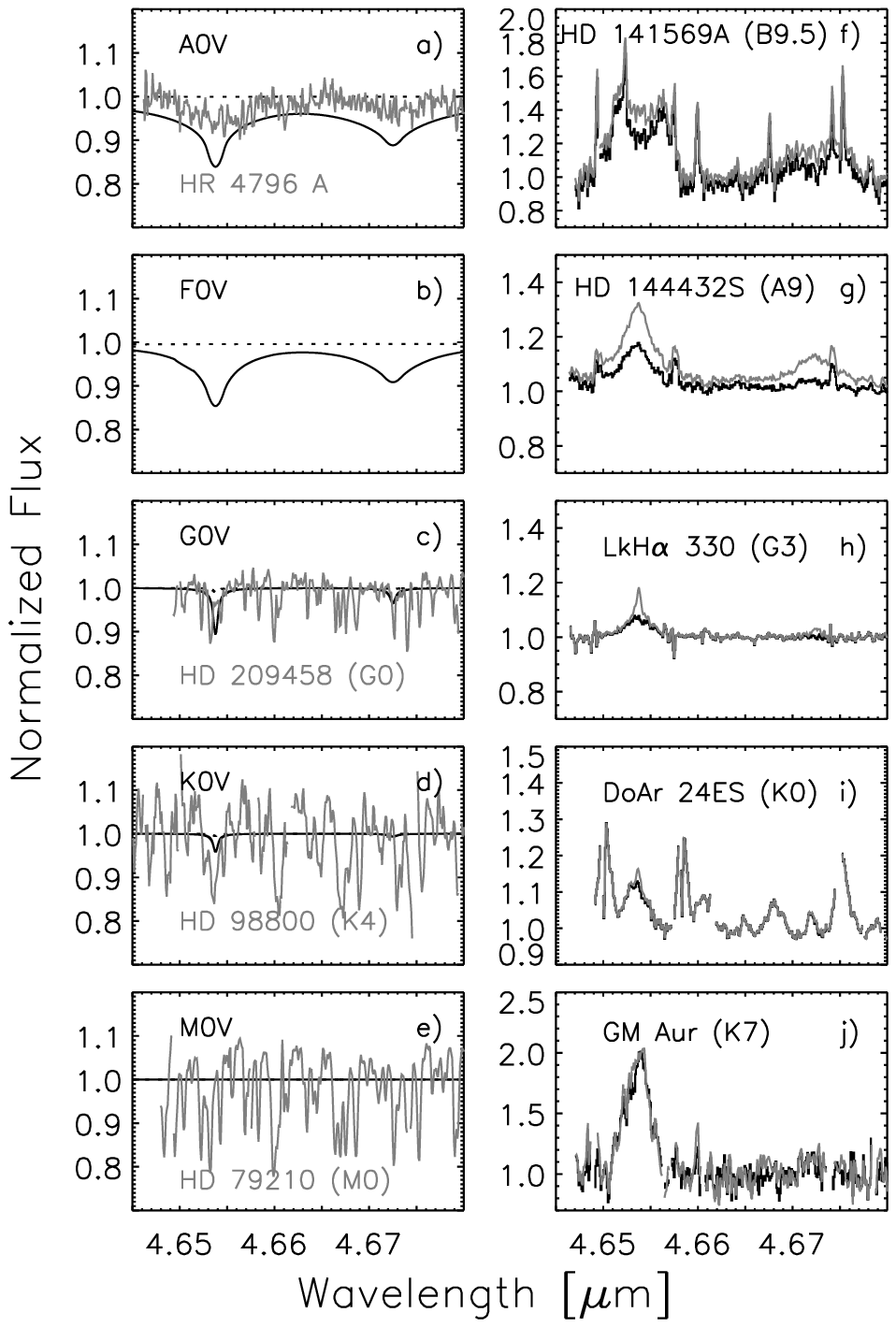}
\caption[something here]{Panels a)-e) show synthetic stellar H \textsc{i} absorption spectra without (solid curves) and with (dotted curves) the addition of typical 
veiling from an optically thick disk.  \footnote{A0V and F0V spectra are produced from ATLAS models using the ATLAS, WIDTH and SYNTHE Linux port \citep{Kurucz93, Sbordone04, Sbordone05}.  G0V, K0V and M0V models are produced with the MOOG stellar synthesis code \citep{Sneden73} using MARCS atmospheric models \citep{Gustafsson08}, but show only the contribution from H \textsc{i} lines.  The actual spectra are dominated by molecular absorption, and stellar models in this wavelength range have not been well benchmarked.    All spectra assume a rotational broadening of 10 km s$^{-1}$.  Representative veiling values were derived using RADMC \citep{Dullemond04}, assuming optically thick disks that extend to $T_\mathrm{in} = 1500$ K.}  Panels a), c), d) and e) additionally show observed spectra of unveiled stars, with the K4 and M0 spectra demonstrating that the spectra of cool stars are dominated by a molecular pseudo-continuum.  Panels f)-j) show example spectra (black) and {\it worst-case} Pf$\beta$ flux corrections (i.e., assuming no veiling; gray) for the following targets: f) transitional disk HD 141569 A, g) Herbig Ae/Be star HD 144432 S, h) transitional disk LkH$\alpha$ 330, i) cTTS DoAr 24 ES and j) transitional disk GM Aur. \label{fig:kurucz_plot}}
\end{minipage}
\end{figure*}

Stellar photospheres have H \textsc{i} absorption features that will tend to reduce the amount of observed Pf$\beta$ flux, i.e. Pf$\beta$ fluxes must be increased somewhat to correct for stellar absorption.  The degree of correction depends sensitively on the stellar spectral type and veiling.  Figure \ref{fig:kurucz_plot}, panels a--e, show synthetic spectra and observed spectra for five spectral types (A0V -- M0V), demonstrating that the absorption is most prominent at early spectral types.  However, M-band veiling values for optically thick disks act in the opposite sense, with much higher veiling around early-type stars. Thus optically thick disks have negligible corrections to Pf$\beta$ from underlying stellar absorption.  Synthetic veiled spectra with veiling values typical of optically thick disks ($r=46$, 18, 10, 7 and 3 for spectral types A--M, derived from disk SEDs produced with RADMC; \citealp{Dullemond04}) are shown as dotted lines, and are indistinguishable from straight lines.

Stars with G--M spectral types have M-band spectra that are dominated by a molecular pseudo-continuum, rather than simply H \textsc{i} absorption (see panels c, d and e), which can result in some change to Pf$\beta$ flux due to coincident molecular absorption features near Pf$\beta$.   The equivalent width of the underlying photospheric features drop by a factor of $1+r$ where $r$ is the veiling (defined as the dust continuum flux density divided by the stellar photospheric flux density).  Considering the spectrum of HD 79210 shown in Figure \ref{fig:kurucz_plot}, a reasonable M-band veiling value of $\sim$3 for an optically thick disk reduces the strongest photospheric features to the $\sim7\%$ level; this and their narrow width results in a negligible correction for optically thick disks.

Transitional disks --- disks with inner regions depleted of small dust grains --- have low continuum fluxes, and so could potentially have large relative contributions from the underlying photosphere.  However, we find that such targets (see Figure \ref{fig:kurucz_plot} panels f), h) and j)) often have high line/continuum ratios, likely because in spite of their lowered near-IR continuum fluxes, transitional disks have only slightly lowered accretion rates \citep{Najita07}.  (See extended discussion in Section \ref{section:discussion}).  As examples, corrections to the Pf$\beta$ flux due to photospheric H \textsc{i} from transitional disks HD 141569 A, LkH$\alpha$ 330 and GM Aur, (panels f, h and j) would be only $\sim$ 20, 45 and 3\%, respectively, even in the worst case scenario (no veiling).  Actual corrections for these sources should be a factor of a few lower, due to non-zero M-band veiling \citep{Salyk09}.  The molecular photospheric absorption features might be more problematic, but luckily the spectra can be examined at other wavelengths to determine whether there is likely significant contamination from the underlying photosphere.  With visual examination, we found one target --- DoAr 21 --- which shows both low Pf$\beta$ equivalent width and low veiling, and in this case the molecular absorption features in its stellar photosphere have a non-trivial effect on the determination of the disk Pf$\beta$ line flux.  We describe this example in detail in an Appendix.

The largest possible correction would be required for a naked A or F star with a low Pf$\beta$ line/continuum ratio (see, e.g., panel g), perhaps doubling the ratio of true to observed Pf$\beta$ emission flux.  Such a target would be difficult to identify because its relatively featureless M-band spectrum looks the same whether it is naked or veiled.  To our knowledge, there are no known diskless targets in our sample that also show detectable Pf$\beta$ emission.  Of course, stellar Pf$\beta$ absorption could erase signatures of weak Pf$\beta$ emission from weakly-accreting disks; however, we have no good way to pinpoint such targets.  A possible example of such a target could be SR 21 (see Figure \ref{fig:nirspec_profiles}) a G2.5 star with a transitional disk, which has CO gas at $\sim$7 AU but no measurable accretion \citep{Pontoppidan08}.

\section{Pf$\beta$ LUMINOSITY AND ACCRETION RATES}
\subsection{Correlation between Pf$\beta$ Luminosity and Accretion Luminosity}
\label{section:correlation}
\begin{figure}
\epsscale{.9}
\plotone{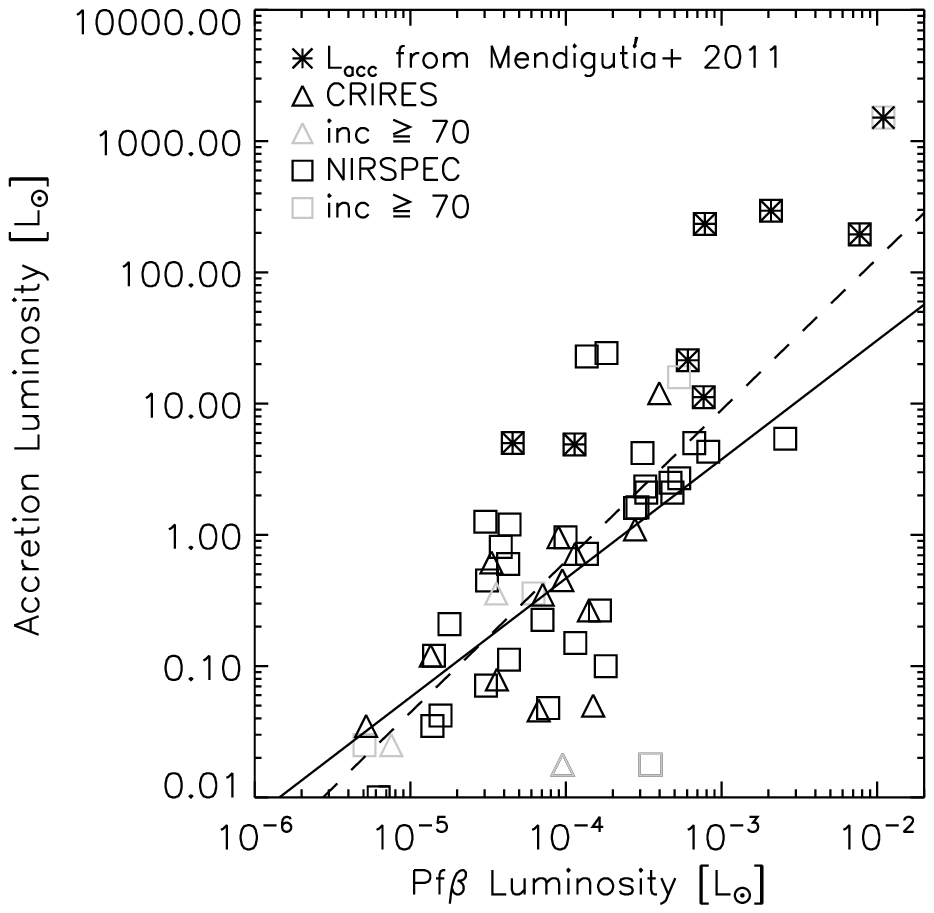}
\caption[place holder]{Correlation between literature accretion luminosity (see Table \ref{table:params}) and Pf$\beta$ luminosity (this work).   If Pf$\beta$ is measured by both NIRSPEC and CRIRES, we show two symbols (squares and triangles, respectively) on the plot. Sources with high inclination or otherwise unreliable accretion rates (in gray) are excluded from the correlation analysis. The solid line shows the best linear fit to the data excluding points from Mendigut\'{i}a et al. (2011) (Equation \ref{eqn:corr}).  The dashed line shows the best linear fit including points where $L_\mathrm{acc}$ is derived from Mendigut\'ia et al. (2011).}
\label{fig:corr_plot}
\end{figure}

In Figure \ref{fig:corr_plot}, we show the correlation between our measured Pf$\beta$ line luminosities and previously-measured accretion luminosities, listed in Table \ref{table:acc_lum}. We chose to maximize the number of sources in this correlation and, therefore, the literature accretion luminosities are derived from a variety of observational methods.  The majority are derived from models and observations of the UV excess continuum, while others are derived somewhat less directly from UV data, utilizing a relationship between mass accretion and the Balmer discontinuity \citep{Muzerolle04} or approximating the UV excess using photometry.  About a third of the accretion rates are derived from empirical relationships between H \textsc{i} emission lines (Br$\gamma$ or Pa$\beta$) and accretion luminosity.  Although the Br$\gamma$ and Pa$\beta$-derived accretion rates are themselves an indirect measure of accretion rate, we find no significant biases between these and the UV-derived accretion luminosities in Figure \ref{fig:corr_plot}.

Because the accretion rates of high-inclination sources can be unreliable, we remove these from our correlation.  We also omit HL Tau from the correlation; although its disk appears to be viewed at moderate inclination \citep{Kwon11}, its high level of veiling makes literature accretion rates very uncertain \citep[e.g.,][]{White01}.  

We find that the correlation between Pf$\beta$ line luminosity and accretion luminosity is heavily influenced by targets with high accretion luminosities derived by \citet{Mendigutia11}. \citet{Mendigutia11} derived accretion rates in 38 HAeBe stars by estimating the Balmer excess using U- and B-band photometry compared with accretion shock models.  Since accretion rates correlate with stellar mass, these accretion rates populate the upper right corner of Figure \ref{fig:corr_plot}.  If these points are included in the fit, we find that our correlation then poorly reproduces previously-measured accretion rates for low-mass stars.  Indeed, \citet{Mendigutia11} note a change in the slope of the correlation between accretion rate and stellar mass between low-mass and high-mass stars, and find larger accretion rates than those determined from Br$\gamma$ emission lines \citep{GarciaLopez06}, implying that correlations between accretion tracers and accretion rates may not extend linearly to higher mass stars.  However, UV excesses are significantly more difficult to measure around high-mass stars, and additional study is warranted to understand whether the emission line-accretion luminosity relationship needs to be modified at higher stellar masses.  Therefore, in this study, we omit the accretion luminosities derived by \citet{Mendigutia11} from our analysis.  If the results of \citet{Mendigutia11} are correct, our study will tend to underestimate accretion luminosities for $M_\star \gtrsim 3 M_\odot$.

Excluding the accretion luminosities in \citet{Mendigutia11}, we find the following relationship between Pf$\beta$ luminosity and previously-measured accretion luminosities:
\begin{equation}
\log{L_\mathrm{acc} [L_\odot]}=(0.91\pm0.16) \times \log{ L_\mathrm{Pf\beta} [L_\odot] }+ (3.29\pm0.67). \label{eqn:corr}
\end{equation}

\subsection{New Accretion Rates}
\label{section:accretion}
We use Equation \ref{eqn:corr} to compute accretion luminosities for our entire sample.  These are listed in Table \ref{table:pf}.  An estimate of $\dot{M}$ is derived using the standard relationship between $L_\mathrm{acc}$ and $\dot{M}$:
\begin{equation}
L_\mathrm{acc}= 0.8\, G M_\star \dot{M}/R_\star
\end{equation}
from \citet{Gullbring98}.
$M_\star$ and $R_\star$ are taken from Table \ref{table:params}. If not shown in Table \ref{table:params}, $R_\star$ is derived from $L_\star$ and $T_\star$ assuming $L_\star=4 \pi R_\star^2 \sigma T_\star^4$.  We do not compute $\dot{M}$ if these parameters are not available.  However, $\dot{M}$ could easily be computed at a later time when these parameters are measured. 

\begin{figure}
\epsscale{.9}
\plotone{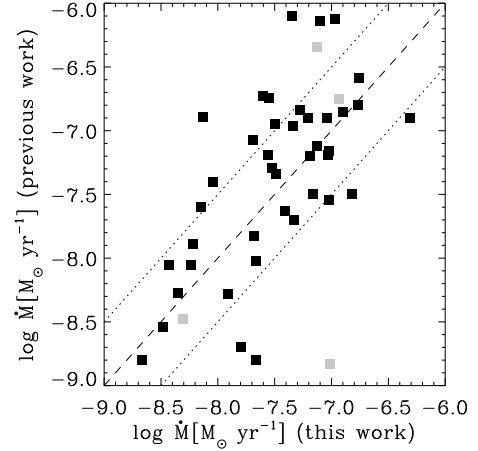}
\caption{Mass accretion rates derived from this work compared with values from the literature.  High-inclination disks are shown plotted in gray.  The dashed line shows a 1:1 correlation and dotted lines show a difference of 0.5 dex.}
\label{fig:mdot_mdot_plot}
\end{figure}

In Figure \ref{fig:mdot_mdot_plot}, we show accretion rates derived here compared to existing values from the literature (Table \ref{table:acc_lum}).  The accretion rates derived here show no systematic bias, and most accretion rates are within 0.5 dex of previous measurements.  The standard deviation of the current and previous measurements is 0.77 dex.  These results are similar to those for other emission line accretion tracers, albeit over a somewhat more limited range of accretion rates. For example, \citet{Herczeg08} quote standard deviations of 1.0, 0.71 and 1.1 dex between UV continuum excess and Ca  \textsc{ii} $\lambda$8542, Ca \textsc{ii} $\lambda8662$ and He \textsc{I} $\lambda 8576$, probing accretion rates as low as $10^{-12} M_\odot \mathrm{yr}^{-1}$.  In addition, \cite{Herczeg08} estimate a factor of 4 (0.6 dex) random error in UV excess measures alone, due to errors in extinction and distance.  Also, accretion rates are known to be variable, with \citet{Nguyen09} finding typical accretion rate variations of 0.35 dex, but variations commonly as high as 0.5 dex.   Therefore, Pf$\beta$ appears to be an accretion tracer on par with existing tracers, with scatter not much larger than what would be expected from accretion variability and errors related to the derivation of UV-based accretion luminosities.

\subsection{About Upper Limits}
Computing upper limits on the Pf$\beta$ line flux, and thus accretion rate, for any given source is not straightforward, and because these are not crucial for the work presented here, we do not make an attempt to compute them.  Complications include the fact that the Pf$\beta$ line shapes are complex, and their widths varied, so the correct choice of assumed lineshape and width is not obvious.  In addition, sources with no detectable Pf$\beta$ emission are often less-heavily veiled sources, whose underlying photospheres must be modeled to derive an accurate Pf$\beta$ flux.  We encourage those interested in an upper limit for a particular source in the NIRSPEC or CRIRES archives to contact the authors of this work to discuss whether reasonable assumptions can be made to derive limits for this source.

Nevertheless, we can make some general observations about how sensitive Pf$\beta$ measurements are to accretion rates.  Assuming the Pf$\beta$ line is a Gaussian with $\sigma_{width}=0.001 \mu\mathrm{m}$ (derived from a fit to AB Aur's Pf$\beta$ emission line), then with a continuum flux density of $F_\nu=1$ Jy, a signal-to-noise ratio (SNR) of 50, a 3$\sigma$ line peak, and for a 1 solar mass, 1 solar radius star, the minimum measurable accretion rate is $\sim4\times10^{-9} M_\odot \mathrm{yr}^{-1}$.  The minimum measurable accretion rate then scales roughly proportional to $\frac{F_\nu}{SNR}$.  For the integration times used for the acquisition of the data used in this work, Pf$\beta$-derived accretion rates do not appear to be as sensitive as previous studies using Br$\gamma$ or Pa$\beta$ \citep[e.g.,][]{Natta06}.   

The lowest detected accretion rates in our sample are for the transitional disks TW Hya and DM Tau, with rates of $2.6\times10^{-9}$ and $4.3\times10^{-9} M_\odot \mathrm{yr}^{-1}$, respectively.  Pf$\beta$ is especially well-suited for measuring small accretion rates from transitional disks, as the dust-cleared inner holes decrease $F_\nu$ and increase the line/continuum ratio (see Section \ref{sec:transitional} for more detail).  However, as the SNR per unit exposure time also decreases as 
$F_\nu$ decreases, sufficient time must be expended to reach a reasonable SNR.  

\section{ANALYSIS AND DISCUSSION}
\label{section:discussion}

\subsection{Utility of Pf$\beta$ and Comparison to Other Accretion Tracers \label{sec:utility}}
In this work, we have introduced the use of H \textsc{i} Pf$\beta$ to measure mass accretion from protoplanetary disks.  Although numerous accretion tracers have been developed prior to this work, the use of Pf$\beta$ offers several advantages, which we outline here.  

One major advantage of using Pf$\beta$ as an accretion tracer is that it comes ``for free'' and contemporaneous with observations of CO fundamental emission.  The NIRSPEC and CRIRES surveys from which we extract the data in this work provide a self-consistent sample of accretion luminosity estimates for \nstars targets, and the NIRSPEC and CRIRES archives likely include many additional protoplanetary disk targets for which accretion luminosities could be estimated.   To our knowledge, this is the first work to make extensive use of M-band spectra of protostars from both NIRSPEC and CRIRES --- data which are all currently available in their respective archives.  The varied and large set of targets resulting from this combined dataset is a great demonstration of the utility of these large archives.  As future work, we plan to extend our analysis to investigate long-term ($\sim$year timescales) accretion variability for different classes of disks. 

The simultaneous observation of CO and Pf$\beta$ also allows for a comparison between the accretion flow and molecular disk gas at distances of $\sim$ 0.1--1 AU from the star \citep[e.g.,][]{Pontoppidan11b, Salyk11b}, which is traced by CO fundamental emission.  The relationship between these two disk components is being investigated, e.g., by \citet{Brown13}.  The contemporaneous nature of these measurements also allows for simultaneous measurements of disk gas and accretion rate as a function of time, although it should be cautioned that there can be changes in emission line fluxes without corresponding changes in accretion rate \citep[e.g.,][]{Gahm08}.   The tidal influence of protoplanets is predicted to cause CO emission line variability \citep{Regaly11}, but understanding the influence of proto-planets requires ruling out other possible sources of line variability, such as changes in accretion rate or accretion geometry.  

\begin{figure}
\includegraphics[scale=0.7]{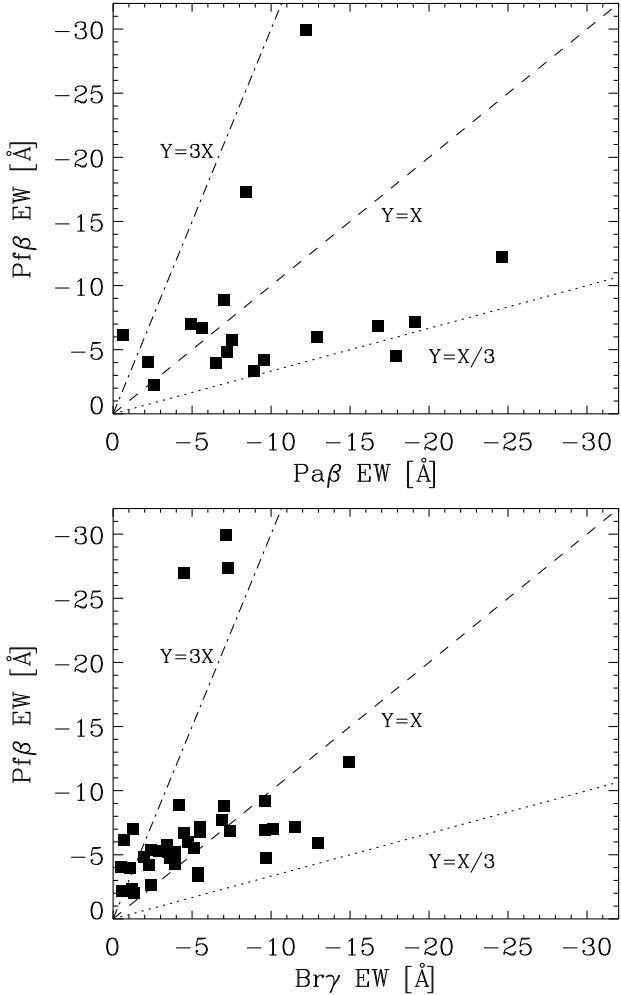}
\caption{EW of Pf$\beta$ compared to those of Br$\gamma$ (bottom) and Pa$\beta$ (top).  The outliers with highest Pf$\beta$ EW in the bottom plot are transitional disks HD 141569 A, GM Aur and TW Hya.  
\label{fig:brgamma_plot}}
\end{figure}

Pf$\beta$ (n=7$\rightarrow$5) is in many ways similar to Br$\gamma$ (n=7$\rightarrow$4; 2.160 $\mu$m), originating from the same upper level energy and appearing in the infrared.  In Figure \ref{fig:brgamma_plot}, we compare Pf$\beta$ EW with literature values of EW(Br$\gamma$) as well as EW(Pa$\beta$) (see Table \ref{table:pf}).  Typical values of Pf$\beta$ equivalent widths are between 1/3 and 3 $\times$ EW(Br$\gamma$) or EW(Pa$\beta$), with EW(Pf$\beta$) typically being similar to EW(Br$\gamma$) but somewhat smaller than EW(Pa$\beta$).  The lack of a strong correlation between tracers likely reflects the varied temperatures of the accretion column, the underlying photosphere, and the continuum veiling.  \citet{Muzerolle98c}, \citet{Calvet04} and \citet{Donehew11} all find correlations between accretion luminosity and Br$\gamma$ luminosity very similar to our Equation \ref{eqn:corr}. Equation \ref{eqn:corr} combined with the relationships in \citet{Muzerolle98c} and \citet{Donehew11} predict somewhat higher EW(Pf$\beta$) than EW(Br$\gamma$), while the relationship in \citet{Calvet04} predicts somewhat lower EW(Pf$\beta$) as compared to EW(Br$\gamma$) (with differences less than $\sim0.2$ dex).  Lineshapes of the two accretion tracers appear to be quite similar --- typically single-peaked with line widths near $\sim$ 200 km s$^{-1}$ \citep{GarciaLopez06}.  HD 141569 A, on the other hand, is an example of a disk with both double-peaked Br$\gamma$ and Pf$\beta$ emission lines.  Interestingly, however, several targets for which Pf$\beta$ is clearly in emission show no Br$\gamma$, or Br$\gamma$ in absorption --- for example, HD 142527, HD 142666,  T CrA and TY CrA.  Thus,  Pf$\beta$ may be a more robust accretion tracer in targets with low near-IR veiling.   

Although equivalent widths are similar, Pf$\beta$ offers some distinct advantages over Br$\gamma$ and Pa$\beta$.  One advantage is the relatively higher continuum veiling in the M band as compared to shorter wavelengths.   Since the photospheric emission is a smaller fraction of the total flux at 5 $\mu$m than it is at shorter wavelengths, Pf$\beta$ suffers less contamination and uncertainty from the underlying photospheric H \textsc{i} absorption (as discussed in detail in Section \ref{sec:photosphere}).  Being at longer wavelengths also makes Pf$\beta$ less sensitive to extinction.  Using the reddening law of \citet{Cardelli89}, $A_\mathrm{Br\gamma}/A_V = 0.12$ while  $A_\mathrm{Pf\beta}/A_V = 0.03$, a factor of 4 difference in {\it magnitudes} of correction.  At $A_V\sim30$, for example, this results in a flux correction factor of $\sim$ 28 for Br$\gamma$ but only $\sim$3 for Pf$\beta$.  Finally, Pf$\beta$ equivalent widths are quite high for accreting disks with low continuum veiling; thus, Pf$\beta$ is an excellent tracer for low accretion rates in disks with inner regions depleted in small dust grains (discussed further in Section \ref{sec:transitional}).  One disadvantage of Pf$\beta$ as an accretion tracer is the systematic errors that are introduced by complex CO emission/absorption spectra, which can result in up to a factor of a few uncertainty in flux in extreme cases (as in Figure \ref{fig:hltau}).  Another is that the veiling is difficult to determine empirically in earlier-type stars with few photospheric features, and so the importance of H \textsc{i} stellar photospheric absorption is difficult to assess in these stars (see Section \ref{sec:photosphere}).

\subsection{Accretion in Transitional Disks}
\label{sec:transitional}
\subsubsection{Equivalent Widths and the Identification of Transitional Disks}
\begin{figure}
\epsscale{.9}
\plotone{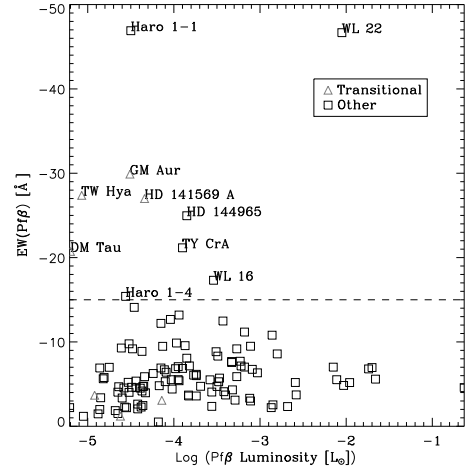}
\caption{Pf$\beta$ equivalent width (EW) plotted against Pf$\beta$ line luminosity.  Note that EW is not a good predictor for Pf$\beta$ luminosity (or therefore accretion luminosity).  The dashed line marks an EW of $-15$ \AA; several transitional disks have extreme values of Pf$\beta$ EW.}
\label{fig:ew_plot}
\end{figure}

In Figure \ref{fig:ew_plot} we show Pf$\beta$ equivalent width (EW) vs.\ Pf $\beta$ luminosity.  There is no significant correlation between these two variables and thus Pf$\beta$ EW is not a good predictor of accretion luminosity.  However, we note that several transitional disks have notably high Pf$\beta$ EW's (between $-20$ and $-30$ \AA). Since the Pf$\beta$ luminosities of these disks are not higher than average, the high EW's instead reflect the fact that these disks have reduced continuum flux levels in the near-IR.  This suggests that Pf$\beta$ may be a good tracer for some accreting transitional disks, and a means to detect these disks with a single spectrum --- one simple, robust measurement that requires no knowledge of the stellar parameters or absolute flux level.  We arbitrarily label all disks with Pf$\beta$ EW $<-15$ and suggest that these targets may have depletions of dust in their inner regions.  Aside from the transitional disks, which are known to have inner disk dust depletions, we also note that TY CrA is a tertiary system with an eclipsing binary \citep[e.g.,][]{Vaz01} and that Haro 1-1 is an anomalously fast rotator \citep{Shevchenko98}. To our knowledge, no unique properties have been discussed for the other disks with high Pf$\beta$ EW.

It is interesting to note that Pf$\beta$ EW's are not high for all transitional disks.  In our sample, LkH$\alpha$ 330, HD 135344 B, IRS 48 and DoAr 44 have typical or even slightly low Pf$\beta$ EW's.  LkH$\alpha$ 330 and HD 135344 B appear to have relatively low gas/dust ratios as compared to other transitional disks \citep{Salyk09}.  DoAr 44 might better be considered a pre-transitional disk \citep{Andrews11} --- a disk with an inner clearing but relatively high near-IR flux, that may be consistent with a small gap rather than a large clearing \citep{Espaillat07}.  And IRS 48 has a $\sim$ 30 AU hole in its gas distribution \citep{Brown12}. Therefore, high Pf$\beta$ EW may be an indicator of only those transitional disks that have significantly cleared their inner regions of small dust grains but not gas.  Thus, a measurement of Pf$\beta$ EW might shed light on the process causing the inner disk depletion for any given disk, as different clearing scenarios predict different gas/dust ratios.

\subsubsection{Transitional Disks in Comparison to Other Disks}
\begin{figure}
\epsscale{1.}
\plotone{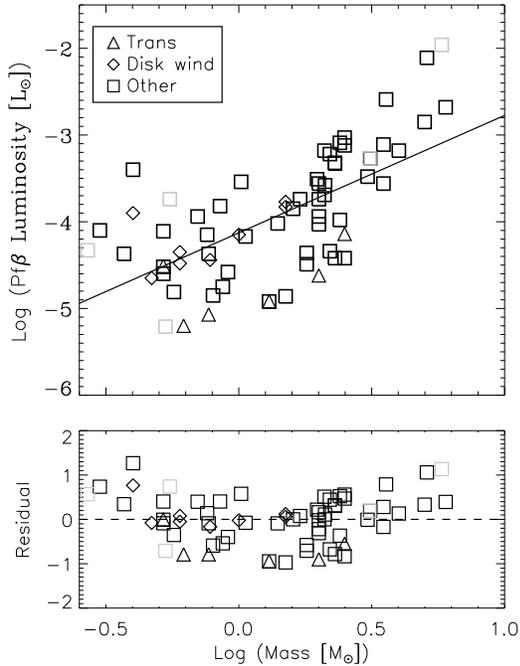}
\caption{The top panel shows the log of Pf$\beta$ line luminosity plotted against the log of stellar mass and best-fit correlation.
The lower panel shows the fit residuals.  Triangles and diamonds are transitional and disk wind targets, respectively; squares are other targets.  Gray squares represent highly-inclined disks.  }
\label{fig:pf_mass_plot}
\end{figure}

Figure \ref{fig:pf_mass_plot} shows the log of Pf$\beta$ luminosity as a function of the log of stellar mass.  Accretion luminosity and accretion rate are known to scale with stellar mass, and these results are no exception.  Plotting Pf$\beta$ luminosity as a function of stellar mass allows us to investigate potential outliers with high or low accretion rates.

Transitional disks have on average lower Pf$\beta$ luminosities, at a marginally statistically-significant level, with a mean log residual Pf$\beta$ luminosity of $-0.7 \pm 0.3$. A low Pf$\beta$ luminosity could represent a low accretion luminosity, or could result if the transitional disks have normal accretion luminosities but an anomalously low Pf$\beta$ luminosity/accretion luminosity ratio.  However, we have investigated the Pf$\beta$ luminosity/accretion luminosity ratio for these targets and find that 4 of 5 transitional disks with literature accretion luminosities (DM Tau, GM Aur, HD 141569 A and TW Hya)  have Pf$\beta$ luminosities even higher than would be predicted by Equation \ref{eqn:corr} (one, HD 135344 B, has a slightly but not significantly lower Pf$\beta$ luminosity).  Therefore, this study is in closer agreement with \citet{Najita07} and \citet{Espaillat12} who find slightly lower than average accretion rates for transitional disks, than with contrasting studies showing no difference between the two types of disks \citep{Fang09, Sicilia-Aguilar10}.  As suggested by \citep{Sicilia-Aguilar10}, the contrasting study results may be due to the different physical nature of transitional disks in different regions.  However, it should be cautioned that the high Pf$\beta$ EW's for transitional disks could produce an observational bias in our work, making it easier to detect a low accretion rate around a transitional, rather than classical, disk.  Thus, our sample may be missing a number of classical disks with low accretion rates.

Since our sample derives from a large number of clusters of different ages, it is also worth asking whether we might measure spuriously low Pf$\beta$ luminosities for transitional disks because they are derived from older clusters, and there is evidence that accretion rates decrease with stellar age \citep{Hartmann98}.  However, we find   
that the transitional disks are derived from a large number of the clusters sampled in this work, and that regardless of the cluster under consideration, transitional disks lie at the low end of the range of measured Pf$\beta$ luminosities.  In addition, the transitional disks are derived primarily from clusters with ages $\lesssim 3$ Myr (with the exception of TW Hya), and so are not biased towards older ages.  

\subsection{Accretion in Disk Wind Sources}
Disk wind targets are a newly identified subset of disks defined by their single-peaked near-IR emission line flux and spectro-astrometric profiles, which may be evidence for the presence of a slow-velocity disk wind \citep{Pontoppidan11b, Bast11}.  
By comparing average accretion luminosities for disk wind and normal disk targets, \citet{Bast11} noted that disk wind targets may have higher than average accretion luminosities.  However, Figure \ref{fig:pf_mass_plot} shows that the Pf$\beta$ luminosities for these targets are actually typical for sources in our sample, implying no higher accretion luminosities.  These targets also have typical ratios of Pf$\beta$ luminosity to derived accretion luminosity.  And, the disk wind sources derive from a large number of clusters, implying no obvious age-related biases.

How can these two results be reconciled?  We believe there are a few possible explanations.  Firstly, the disk wind sources in \citet{Bast11} have somewhat higher stellar masses than their comparative sample of ``normal'' disks.  Secondly, the disks in the comparative sample in \citet{Bast11} have lower accretion luminosities at a given mass
than disks in our sample.  Since our full sample has not been ``vetted'' for disk wind targets, it is possible that our full sample includes unidentified disk wind targets that increase the average, or our sample may simply be biased towards higher accretion rates.  Unfortunately, we cannot use the criteria in \citet{Bast11} to classify all of our targets because of the relatively lower spectral resolution of NIRSPEC as compared to CRIRES.  

The accretion rates for the disk wind targets, derived from this work, range from 6$\times10^{-9}$ to 2$\times10^{-7}M_\odot \mathrm{yr}^{-1}$, with most accretion rates in the range of $10^{-8}-10^{-7}M_\odot \mathrm{yr}^{-1}$.  Thus, the disk wind sources are likely capable of supporting mass-loss rates of $10^{-10}-10^{-9}M_\odot \mathrm{yr}^{-1}$, and in some cases $10^{-8} M_\odot \mathrm{yr}^{-1}$, as suggested by \citet{Pontoppidan11b}.

Although we find typical accretion rates for disk wind sources, it should be noted that one of the primary characteristics of disk wind targets is their high veiling at optical through IR wavelengths \citep[e.g.,][]{Gahm08}.  The seemingly contradictory observation of high veiling but normal accretion rates may be reconciled by the results of \citet{Gahm08}, who find that the optical veiling in such targets is not a good measure of disk accretion, as the derived optical veiling is affected by infilling of photospheric lines from line emission.

\subsection{Accretion Rates for Embedded Disks}
\label{sec:embedded}
\subsubsection{Classification of Evolutionary Stage}

The process of star and planet formation is often deconstructed into four discrete evolutionary stages.  The targets in our sample represent primarily, if not entirely, sources in stages I and II, where stage I sources are embedded protostars with still-collapsing envelopes, while stage II sources are revealed stars with circumstellar disks and little or no remaining envelope.  For this work, the evolutionary class was determined primarily through examination of the M-band spectra themselves, with stage I disks defined as those showing CO ice absorption, and stage II disks those without.   This offers the significant advantage that our entire sample can be easily classified using our own dataset, with a single consistent criterion.

However, it is a known issue that edge-on stage II disks can masquerade as evolutionarily-younger systems, especially with the use of spectral slope-based classification schemes (see, for example, the detailed discussion in \citealp{Evans09}).  Our classification scheme is not immune to this issue, as CO ices in edge-on disks can in principle produce absorption features. However, in practice, the production of strong ice absorption features requires a very specific geometry, with an inclination close to the opening angle of the flaring disk.  Additionally, the upper layers of the disk that intercept the starlight are often too warm to contain significant quantities of CO ice \citep{Pontoppidan05}.  Thus, it is perhaps not surprising that our classification scheme correctly identifies some high-inclination disks, including T Tau S and VV Ser, as stage II rather than stage I sources.

Nevertheless, we have made an effort to confirm stage I classifications whenever possible using additional diagnostics in the literature, including the use of spectral indices, bolometric temperatures, the presence of infall or outflow signatures in spectral lines, and the presence of extended envelopes as seen in millimeter-wave maps \citep[e.g.][]{Boogert02, White04, Doppmann05, Evans09, vanKempen09, Jorgensen09, Herczeg11}.  Our classifications are listed in Table \ref{table:params}, along with accompanying notes for sources with ambiguous properties.

\subsubsection{Accretion Rates for Stage I Disks, and Comparison with Stage II Disks}
Since stars are believed to accumulate most of their mass during their embedded stages, accretion rates for stage I disks were initially predicted to be significantly larger than for stage II disks.   Smooth collapse models \citep{Shu77} predicted stage I accretion rates similar to simple estimates --- for example, accretion rates of order $\gtrsim 10^{-6}M_\odot \mathrm{yr}^{-1}$ are required to build a solar mass star over a 
$\sim10^6$ year timescale. Accretion rates in stage II disks around solar-mass stars are typically near $\sim 10^{-8}M_\odot \mathrm{yr}^{-1}$ (see Table \ref{table:acc_lum} and references therein), and thus stage I accretion rates would be predicted to be at least two orders of magnitude higher.

While there is at least some evidence for greater accretion activity in stage I sources \citep[e.g.][]{Doppmann05}, measured accretion rates in stage I disks around low-mass stars are found to be near $10^{-8}M_\odot \mathrm{yr}^{-1}$ \citep[e.g.][]{White04, White07} --- two orders of magnitude lower than predicted.  This discrepancy is related to the well-known ``luminosity problem'' in star formation \citep{Kenyon90}, wherein measured bolometric luminosities (which derive primarily from the energy of gravitational contraction, and are therefore related to mass accretion) are lower than predicted from steady growth models.  Instead, evidence is growing \citep[e.g.][]{Dunham08, Enoch09, Evans09} that accretion may need to be episodic --- primarily quiescent, with brief phases of intense accretion.  

While direct measurements of accretion rates in stage I disks appear to support this hypothesis, such studies inevitably suffer from a bias towards more revealed disks, as the most embedded protostars are heavily extincted, and cannot be easily studied with either optical or near-IR accretion tracers.  Although it is difficult to quantify the many sample-selection biases here, extinction remains small in the M band even for high $A_V$, and our sample includes several heavily embedded protostars. As a point of comparison, the sample of young protostars in \citet{White04} all have $A_V<20$, while 10 of 20 stage I disks in our study have $A_V\geq20$.  Thus, our study provides insight about accretion rates for the most embedded, and perhaps evolutionarily youngest, disks.

\begin{figure}
\epsscale{1}
\plotone{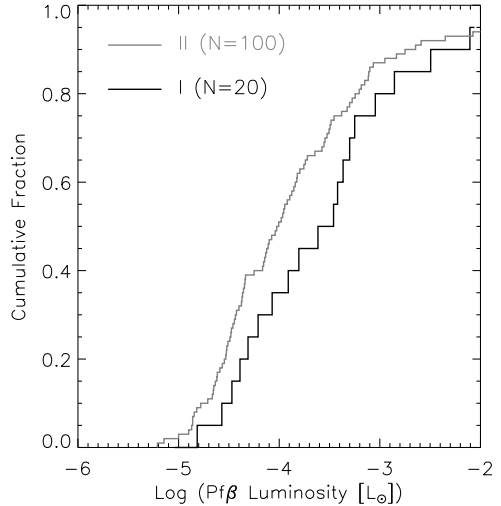}
\caption{The cumulative distribution function of observed Pf$\beta$ accretion luminosities for stage I and stage II disks.}
\label{fig:class_edf}
\end{figure}

Figure \ref{fig:class_edf} shows the empirical cumulative distributions of calculated Pf$\beta$ luminosities for stage I and stage II disks in our sample. There are small differences between the distributions of Pf$\beta$ line luminosities for the two samples, with slightly higher Pf$\beta$ luminosities in Stage II disks.   Stage I disks have a mean $\log(L_{\mathrm{Pf}\beta})$ of $-3.7$ and a standard deviation of $0.8$, while stage II disks have a mean $\log(L_{\mathrm{Pf}\beta})-3.8$ with a standard deviation of 0.9.  However, a two-sample Kolmogorov-Smirnov test produces an associated probability of 30\%.  Therefore, the hypothesis that the stage I and stage II disks draw from the same distribution of accretion luminosities cannot be rejected, i.e., differences between the two distributions are not statistically significant.  

Unfortunately, accretion {\it rate} distributions cannot be compared directly, as the majority of the embedded disks in our sample do not have measured stellar masses or radii due to the relative difficulty of obtaining optical spectra for highly embedded stars.  However, if average stellar masses and radii are similar for the two samples our results suggest that accretion rates are quite similar for stage I and stage II disks.  Spectroscopic analyses of stage I protostars have suggested that stage I and stage II stars have similar average masses, while stellar radii are no more than a factor of $\sim2$ larger for stage I stars \citep{White04}.  For the small number (four) of stage I disks with measured stellar masses in our sample, the average stellar mass is $1.7 M_\odot$, as compared to an average of $1.8M_\odot$ for our stage II sample.  Assuming at most a factor of two difference in radius between stage I and II stars, and assuming no signficantly different selection biases for our two subsamples, our results suggest that stage I accretion rates are at most factors of a few higher, on average, than they are for stage II disks. 

We can directly measure the accretion rates of four stage I targets in our sample.  These accretion rates range from $1\times10^{-8}$ to $4\times10^{-7} M_\odot \mathrm{yr}^{-1}$.  Thus, these targets have an accretion rate spread of more than an order of magnitude, spanning from a value typical for low-mass stars to something close to the $10^{-6} M_\odot \mathrm{yr}^{-1}$ required to build a solar mass star in 1 Myr.  If we divide the currently-measured stellar mass by $\dot{M}$, we can obtain a lower limit to the time required to form the star assuming a steady state accretion rate.  We find timescales of 0.7, 6, 54 and 67 Myr for the four stage I sources --- longer than estimated stage I lifetimes \citep{Evans09}.  Therefore, our data suggest that these sources must at times have had significantly higher accretion rates than currently observed.  

Making some simple assumptions, we can also investigate accretion rates for the full stage I sample.  For example, assuming $M_\star=1 M_\odot$ and $R_\star=2R_\odot$ for the 16 stage I stars without measured stellar parameters, we find a mean and standard deviation of $\log(\dot{M})=-7.1\pm{0.7}$.  This value is at least an order of magnitude lower than the predictions from steady accretion models.  Yet, with these assumptions, our sample does include two sources with accretion rates of $\sim2\times10^{-6} M_\odot \mathrm{yr}^{-1}$.  If our sample is representative of {\it all} stage I sources and homogeneous except for random changes in accretion rate due to episodic accretion, and taking into account a detection fraction of 60\% (see Section \ref{sec:characteristics}), this would imply that the stage I sources spend $\sim6\%$ of their lifetime with $\dot{M}>10^{-6} M_\odot \mathrm{yr}^{-1}$.  The large number (40\%) of Pf$\beta$ non-detections is further evidence for a wide range of accretion rates in stage I sources, again consistent with an episodic accretion scenario.

It is also interesting to note that no FU Orionis stars in our sample show detectable Pf$\beta$ emission lines.  The explanation for this is not known, but may have to do with the restructuring of the inner disk that occurs during high accretion-rate events \citep[e.g.][]{Zhu09}.  In any case, while these sources may represent protostars with the highest rates of accretion \citep{Hartmann96}, they are necessarily excluded from this analysis due to their lack of Pf$\beta$ emission, and thereby bias the remaining sample towards lower accretion rates.

\section{CONCLUSIONS}
In this work, we have introduced the use of H \textsc{i} Pf$\beta$ as a mass accretion tracer for young stars with protoplanetary disks.  
H \textsc{i} Pf$\beta$ offers several advantages over other accretion tracers, including being readily observed in heavily-extincted disks, requiring minimal correction for photospheric absorption and being commonly observed along with CO fundamental transitions.   Using existing measurements of accretion from the literature, we derive a relationship between Pf$\beta$ line luminosity and accretion luminosity, and show that this relationship can reproduce measured accretion rates with an accuracy similar to that of other commonly-used tracers.  Examining our large sample of accretion rates, we are further able to show that ``disk wind'' sources appear to have normal accretion rates, while our sample of transitional disks have slightly lower than average accretion rates.  

We also examine the accretion rates of the stage I disks in our sample --- a sample that includes significantly more embedded targets than in previous studies of this type.  We find that stage I and stage II disks have statistically indistinguishable Pf$\beta$ luminosities, implying similar accretion rates, and that the accretion rates of stage I disks are too low to build a stellar mass with quiescent accretion.  Our results instead are consistent with both observational and theoretical evidence that stage I objects experience episodic, rather than quiescent, accretion.
 
The calibration of Pf$\beta$ presented here has allowed for a coherent comparison of accretion luminosity in \nstars stars.  While this work has focused exclusively on spectra from the NIRSPEC and CRIRES surveys with which we are associated, the correlation we derive can be extended to a much larger number of protoplanetary disk spectra, including some which may already reside in the NIRSPEC or CRIRES archive.  We have also focused exclusively on disks with clearly detectable Pf$\beta$ emission.  However, with more extensive observations and focused analysis, our correlation can be extended to tenuous disks with low accretion rates --- disks of utmost importance to our understanding of disk dissipation.  Finally, as Pf$\beta$ is observed contemporaneously with CO fundamental emission lines, the observation of Pf$\beta$ can be used to account for the effects of accretion variability in searches for planet-induced CO line variability.  We therefore expect the utility of Pf$\beta$ as an accretion tracer to extend well beyond the work presented here.

\section{ACKNOWLEDGEMENTS}
C.S. thanks Casey Deen and Chris Sneden for providing an introduction to MOOG.  C.S. also acknowledges the financial support of the NOAO Leo Goldberg Fellowship program and the University of Texas at Austin McDonald Observatory Harlan J. Smith Postdoctoral Fellowship.  E.vD. is supported by EU A-ERC grant 291141. Some of the data presented herein were obtained at the W.M. Keck Observatory, which is operated as a scientific partnership among the California Institute of Technology, the University of California and the National Aeronautics and Space Administration. The Observatory was made possible by the generous financial support of the W.M. Keck Foundation. This publication also makes use of data products from the Wide-field Infrared Survey Explorer, which is a joint project of the University of California, Los Angeles, and the Jet Propulsion Laboratory/California Institute of Technology, funded by the National Aeronautics and Space Administration.  

\begin{center}APPENDIX\end{center}
\section*{\uppercase{The special case of DoAr 21 \label{sec:doar21}}}
DoAr 21 is known to be an unusual disk, with little or no reported excess emission below 7 $\mu$m and weak or non-existent accretion, but asymmetric extended emission beyond $\sim100$ AU, and line emission from Polycyclic Aromatic Hydrocarbons (PAH's) and H$_2$ \citep[][and references therein]{Jensen09}.  Observations of DoAr 21 with the VLBA have revealed that DoAr 21 is a 5 mas (0.6 AU) binary \citep{Loinard08}.  Detailed studies of evolved sources like DoAr 21 can provide insight into the process of disk dissipation, as it may represent a case of a disk at the latest stages of dissipation.

\begin{figure*}
\epsscale{.9}
\plotone{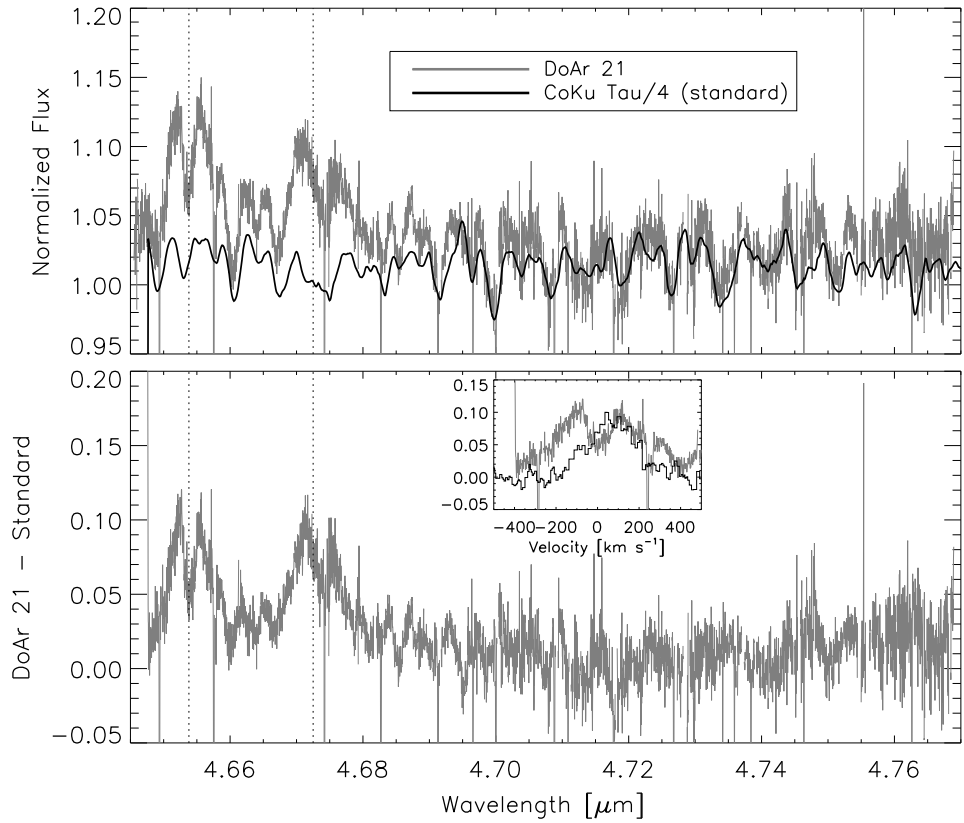}
\caption{Top: Spectrum of DoAr 21 (K1) observed with CRIRES (gray) and a rotationally-broadened, veiled spectrum of CoKu Tau/4 (K3) observed with NIRSPEC (black).  Dotted lines mark the locations of Pf$\beta$ and Hu$\epsilon$.  Bottom: Difference spectrum.  Inset: Difference spectrum centered at 4.6538 $\mu$m (gray) compared with observations of H$\alpha$ from \citet{Jensen09}.}
\label{fig:doar21_plot}
\end{figure*}

DoAr 21 is the only disk in our sample that shows both low but detectable Pf$\beta$ equivalent width, and low veiling, meaning that the underlying photosphere must be accounted for in the calculation of Pf$\beta$ line flux.  In Figure \ref{fig:doar21_plot}, we show the CRIRES spectrum of DoAr 21 (K1) along with a NIRSPEC spectrum of a photospheric template star --- CoKu Tau/4  (K3) --- rotationally broadened and veiled to match the observed spectrum of DoAr 21 longward of 4.68 $\mu$m.  Although DoAr 21 and CoKu Tau/4 do not have the same spectral sub-type, we find only small differences in spectra across a range of observed spectral types (G2--M0), and both are young stars with similarly lowered surface gravities compared to main-sequence stars.  

The bottom of Figure \ref{fig:doar21_plot} shows the difference between the observed and template spectra, which reveals strong and double-peaked Pf$\beta$ and Hu$\epsilon$ emission lines.  As an insert in Figure \ref{fig:doar21_plot}, we compare Pf$\beta$ to an observed spectrum of H$\alpha$ from \citet{Jensen09}. The H$\alpha$ line shown here is in fact the difference between H$\alpha$ lines observed on two different nights, and therefore represents the variable, perhaps flare-related, component to H$\alpha$.  We find that the Pf$\beta$ emission line is broader and more double-peaked than the H$\alpha$ line, suggesting that the Pf$\beta$ may not have a flare origin, but instead have an accretion origin.  (It should be noted, however, that the Hu$\epsilon$ line flux appears unusually high compared to other disks in our sample.)

We also note that the best fit to the observed DoAr 21 spectrum requires a veiling of $\sim$ 2.6, and that the spectrum is inconsistent with zero veiling, consistent with the results of \citet{Salyk09}.   Assuming the continuum flux is emitted entirely by grains at 1500 K with an opacity of 500 cm$^2$g$^{-1}$ (appropriate for sub-micron sized grains), the mass in small grains is $\sim$10$^{19}$ kg, or 
$\sim10^{-4}$ M$_\mathrm{\rightmoon}$.  This is almost certainly a lower limit, as there is likely a range of grain sizes and a range of grain temperatures, with 1500 K being the approximate maximum temperature for solid silicates.  An alternative way to explain finite veiling would be for the stellar disk to contain hotspots with featureless emission spectra.  However, even in an extreme case of 10$^4$ K hotspots, 54\% of the stellar disk would need to be covered by hotspots to explain the observed veiling.  In contrast, results from photometric variability studies \citep{Bouvier95} find hotspot temperatures of no more than a few hundred to a few thousand K hotter than the stellar temperatures, and $0.5-40$\% disk covering fractions.  Thus, DoAr 21 may retain both a gas and dust disk at small radii.

\clearpage
\LongTables
\begin{deluxetable}{lccrc}
\tablecaption{Log of Observations \label{table:obslog}}
\tablehead{\colhead{Star} & \colhead{Date} &\colhead{Exp}&\colhead{Vdop\tablenotemark{a}}&Std\\
& \colhead{[yyyy mmm dd]} &\colhead{[s]}&\colhead{[km s$^{-1}$]}}
\tabletypesize{\footnotesize}
\startdata
NIRSPEC\\
\hline
AA Tau
  &   2003 Nov 03    &   720  &   $-15$ & HR 1620  \\
   &  2004 Dec 27    &   780  &    13  & HR 1620 \\
AB Aur
  &   2001 Jan 30    &   360  &    24  & HR 1620\\
  &   2001 Aug  07    &   180  &   $-25$  & HR 1620\\ 
  &   2002 Jan 03    &   240  &    13  & HR 1620\\
AS 205 N
  &   2002 Apr 21    &   240  &   $-17$  & HR 6175\\
  &   2002 Jul 22    &   480  &    24 & HR 5984 \\
\enddata
\tablenotetext{a}{Earth-induced Doppler shift towards the source position.}
\end{deluxetable}

\begin{deluxetable}{lcccl}
\tablecaption{Photometry \label{table:photometry}}
\tablehead{\colhead{Star} & \colhead{M} & 4.5\tablenotemark{a} & 5.8\tablenotemark{a} & Refs}
\tabletypesize{\footnotesize}
\startdata
              AA Tau &       6.86 &       6.77 &       6.35 &      \ref{Keny95},\ref{Luhm06}\\
              AB Aur &       2.70 &    \nodata &    \nodata &                   \ref{Malf98}\\
            AS 205 N &       4.18 &    \nodata &    \nodata &                   \ref{McCa06}\\
            AS 205 S &       5.31 &    \nodata &    \nodata &                   \ref{McCa06}\\
              AS 209 &    \nodata &    \nodata &       5.05 &                   \ref{Evan03}\\
              BF Ori &    \nodata &    \nodata &    \nodata &                        \nodata\\
              CI Tau &    \nodata &       6.27 &       5.94 &                   \ref{Robi07}\\
                CK 4 &    \nodata &       7.49 &       7.56 &                   \ref{Evan03}\\
\enddata
\tablenotetext{a}{Spitzer-IRAC filters.}
\tablenotetext{b}{Although other photometric measurements are available for this source, we use the spectroscopically-derived measurements from \citet{Goto11} as EX Lup was in outburst during the observations presented here, and thus several magnitudes brighter than in its quiescent state.}
\tablenotetext{c}{Corrected for observed relative fluxes of IRS 44 E and IRS 44 W from \citet{Herczeg11}.}
\tablerefs{
\usecounter{minirefcount}
\mr{Bouvier et al.\ 1992}{Bouv92}
\mr{Carpenter et al.\ 2008}{Carp08}
\mr{Cieza et al.\ 2007}{Ciez07}
\mr{Ducati et al.\ 2002}{Duca02} 
\mr{Evans et al.\ 2003}{Evan03}
\mr{Evans et al.\ 2009}{Evan09}
\mr{Goto et al.\ 2011}{Goto11}
\mr{Hartmann et al.\ 2005}{Hart05}
\mr{Herczeg et al.\ 2011}{Herc11}
\mr{Hillenbrand et al.\ 1992}{Hill92}
\mr{Hughes et al.\ 1994}{Hugh94}
\mr{Kenyon \& Hartmann 1995}{Keny95}
\mr{Lada et al.\ 2006}{Lada06}
\mr{Liseau et al.\ 1992}{Lise92}
\mr{Luhman et al.\ 2006}{Luhm06}
\mr{Malfait et al.\ 1998}{Malf98}
\mr{McCabe et al.\ 2006}{McCa06}
\mr{Mer\'{i}n et al.\ 2004}{Meri04}
\mr{Mer\'{i}n et al.\ 2008}{Meri08}
\mr{Padgett et al.\ 2006}{Padg06}
\mr{Padgett et al.\ 2007}{Padg07}
\mr{Pinte et al.\ 2008}{Pint08}
\mr{Rebull et al.\ 2010}{Rebu10}
\mr{Robitaille et al.\ 2007}{Robi07}
\mr{Schegerer et al.\ 2009}{Sche09}
\mr{Wright et al.\ 2010}{Wrig10}
}
\end{deluxetable}

\clearpage

  \onecolumngrid

\begin{deluxetable*}{lccccccccccc}
\tablecaption{Stellar and disk parameters\label{table:params}}
\tablehead{\colhead{Star} & \colhead{dist} 	&\colhead{SpT}		&\colhead{$T_\star$} 	& \colhead{$L_\star$}	&\colhead{$R_\star$}	&\colhead{$M_\star$}	&\colhead{$A_V$} 	&Refs	& Stage \tablenotemark{a} & Ref & Inst. \tablenotemark{b}}
\startdata
               AA Tau &     140 &      K7 &    4060 &    0.71 &    1.74 &    0.53 &    0.74 &                                                                      \ref{Gul98} &    II & \nodata & N,C   \\
              AB Aur &     140 &      A0 &    9840 &   48 & \nodata &    2.40 &    0.5 &                                                                      \ref{Gar06} &    II &  \nodata & N   \\
            AS 205 N &     160 &      K5 &    4450 &    7.1 & \nodata &    1.50 &    2.9 &                                                                      \ref{Pra03} &    II &  \nodata & N,C   \\
            AS 205 S &     160 &      M3 &    3450 &    2.19 & \nodata &    0.30 &    2.1 &                                                                      \ref{Pra03} &    II &  \nodata & N   \\
              AS 209 &     160 &      K5 &    4395 &    2.5 &    2.40 &    1.40 &    1.15 &                                                                      \ref{Joh00} &    II &  \nodata & C   \\
              BF Ori &     400 &   A5-F6 & \nodata &   27 &    2.70 &    2.10 &    2.4 &                                                          \ref{Blo06},\ref{Lah07} &    II &  \nodata & N   \\
              CI Tau &     140 &      K7 &    4060 &    0.84 & \nodata &    0.77 &    1.8 &                                                                      \ref{Whi01} &    II &  \nodata & N   \\
                CK 4 &     260 &      K3 &    4730 & \nodata & \nodata & \nodata &   12.8 &                                                          \ref{Eva09},\ref{Lah07} &   II\tablenotemark{c} & \ref{Eva09} & N   \\
\enddata
\end{deluxetable*}

\clearpage

\begin{deluxetable}{lrrrrrrc}
\tablecaption{Derived Pf$\beta$ Accretion Parameters; EW's of Other Accretion Tracers \label{table:pf}}
\tablehead{\colhead{Star} & \colhead{EW$_\mathrm{Pf\beta}$}&\colhead{$\log{L_\mathrm{Pf\beta}}$\tablenotemark{a}} & \colhead{$\log{L_\mathrm{acc}}$} & \colhead{$\log{\dot{M}}$\tablenotemark{b}} &\colhead{EW (Br$\gamma$)} & \colhead{EW (Pa$\beta$)} & \colhead{Refs}\\
 &\colhead{[\AA]}& \colhead{[$L_\odot$]} & \colhead{[$L_\odot$]} & \colhead{[$M_\odot\, \mathrm{yr}^{-1}$]}&\colhead{[\AA]}&\colhead{[\AA]}}
\tabletypesize{\footnotesize}
\startdata
               AA Tau &    $-2.21$ &    $-5.21$ &    $-1.43$ &    $-8.31$ &    $-0.59$ &    $-2.58$ &          2\\
              AB Aur &    $-6.75$ &    $-3.09$ &     $0.50$ &   $ -6.90$ &    $-5.50$ &    \nodata &          3\\
            AS 205 N &    $-3.59$ &    $-3.83$ &    $-0.18$ &    $-7.10$ &    \nodata &    \nodata &    \nodata\\
            AS 205 S &    $-5.29$ &    $-4.10$ &    $-0.42$ &   $ -6.68$ &    \nodata &    \nodata &    \nodata\\
              AS 209 &    $-4.41$ &    $-4.02$ &    $-0.35$ &    $-7.52$ &    \nodata &    \nodata &    \nodata\\
              BF Ori &    $-4.74$ &    $-3.69$ &    $-0.05$ &   $ -7.34$ &    $-3.60$ &    \nodata &    \nodata\\
              CI Tau &    $-8.88$ &    $-4.37$ &    $-0.66$ &   $ -7.68$ &    $-4.20$ &    $-7.00$ &          5\\
                CK 4 &   $-12.67$ &    $-4.04$ &    $-0.37$ &    \nodata &    \nodata &    \nodata &    \nodata\\
\enddata
\tablenotetext{a}{Adjusted for photometry and extinction, as described in text.  Assumes distances in Table \ref{table:params}.  If both NIRSPEC and CRIRES data were available, the average value is shown.}
\tablenotetext{b}{Calculated assuming $L_\mathrm{acc}= 0.8\, G M_\star \dot{M}/R_\star$ \citep{Gullbring98}.  If not shown in Table \ref{table:params}, $R_\star$ is derived from $L_\star$ and $T_\star$ assuming $L_\star=4 \pi R_\star^2 \sigma T_\star^4$.  Missing data indicate that either the stellar radius or mass is not available.}
\tablerefs{(1) Donehew  \& Brittain 2011 (2) Folha et al.\ 2001 (3) Garcia Lopez et al.\ 2006 (4) Muzerolle et al. 1998 (5) Natta et al.\ 2006}

\end{deluxetable}

\clearpage

\begin{deluxetable}{lccccc}
\tablecaption{Literature Accretion Luminosities and Mass Accretion Rates \label{table:acc_lum}}
\tablehead{\colhead{Star} & \colhead{L$_\mathrm{acc}$}&\colhead{Ref}&\colhead{Method \tablenotemark{a}}&$\log{\dot{M}}$&\colhead{Ref}\\
&\colhead{[L$_\odot$]}&&&\colhead{[$M_\odot \mathrm{yr}^{-1}$]}\\
}
\startdata
          AA Tau  &     0.025  &       \ref{Gullb98} &          UV Exc &    $-8.48$ &        \ref{Gullb98}\\
         AB Aur  &     4.3  &       \ref{Garci06} &     Br $\gamma$ &    $-6.85$ &        \ref{Garci06}\\
       AS 205 N  &   \nodata  &             \nodata &         \nodata &    $-6.14$ &        \ref{Eisne05}\\
       AS 205 S  &   \nodata  &             \nodata &         \nodata & \nodata &              \nodata\\
         AS 209 &      0.452\tablenotemark{b}&        \ref{Valen93} &          UV Exc &    $-7.29$ &        \ref{Valen93}\\
         BF Ori  &   \nodata  &             \nodata &         \nodata &    $-6.96$ &        \ref{Blond06}\\
         CI Tau  &     0.112  &       \ref{Valen93} &          UV Exc &    $-7.83$ &        \ref{Valen93}\\
           CK 4  &   \nodata  &             \nodata &         \nodata & \nodata &              \nodata\\
\enddata
\tablenotetext{a}{Description of abbreviations are as follows.  UV Exc. = measurement of UV excess emission spectrum; Br $\gamma$ = use of Br $\gamma$-accretion luminosity relationship; Pa $\beta$ = use of Pa $\beta$-accretion luminosity relationship; U-Band Phot. = use of optical U-band photometric data; Balmer depth = based on relationship between the mass accretion rate and the excess in the Balmer discontinuity derived by \citet{Muzerolle04}}
\tablenotetext{b}{This and all other Taurus accretion luminosities from \citet{Valenti93} scaled to 140 pc distance.}
\tablerefs{\usecounter{minirefcount}
\mr{Andrews \& Williams 2007}{Andre07} 
\mr{Andrews et al.\ 2011}{Andre11}
\mr{Aspin et al.\ 2010}{Aspin10_mr}
\mr{Blondel et al.\ 2006}{Blond06}
\mr{Calvet et al.\ 2004}{Calve04}
\mr{Donehew \& Brittain 2011}{Doneh11}	
\mr{Eisner et al.\ 2005}{Eisne05}
\mr{Fern{\'a}ndez et al.\ 1995}{Ferna95} 
\mr{Garcia Lopez et al.\ 2006}{Garci06}
\mr{Gullbring et al.\ 1998}{Gullb98}
\mr{Hartmann et al.\ 1998}{Hartm98}  
\mr{Herczeg et al.\ 2008}{Hercz08}
\mr{Muzerolle et al.\ 1998}{Muzer98}
\mr{Mendigut\'{i}a et al.\ 2011}{Mendi11}
\mr{Natta et al.\ 2006}{Natta06_mr}
\mr{Nisini et al.\ 2005}{Nisin05}
\mr{Prato et al.\ 2009}{Prato09}
\mr{Valenti et al.\ 1993}{Valen93}
\mr{White \& Ghez 2001}{White01}
\mr{White \& Hillenbrand 2004}{White04_mr}   
     }
\end{deluxetable}

%from Bouvier et al. PPV
%"current studies have almost always been limited to H? Ð one that rarely shows an IPC profile (Edwards et al., 1994; Reipurth et al., 1996), is vulnerable to contamination by outflows (e.g., Alencar et al., 2005) and may be signif- icantly spatially extended (Takami et al., 2003). "

%from Bouvier et al. PPV
%The implications of the dynamical nature of magneto- spheric accretion in CTTSs are plentiful and remain to be fully explored. They range from the evolution of stellar angular momentum during the pre-main sequence phase (e.g., Agapitou and Papaloizou, 2000), the origin of in- flow/ouflow short term variability (e.g., Woitas et al., 2002; Lopez-Martin et al., 2003), the modeling of the near in- frared veiling of CTTSs and of its variations, both of which will be affected by a non planar and time variable inner disk structure (e.g., Carpenter et al., 2001; Eiroa et al., 2002), and possibly the halting of planet migration close to the star (Lin et al., 1996).

\end{document}